\documentclass[usenatbib]{mnras}

\usepackage{newtxtext,newtxmath}
\usepackage{verbatim}
\usepackage[T1]{fontenc}
\usepackage{ae,aecompl}
\usepackage{tabularx}

\usepackage{graphicx}	% Including figure files
\usepackage{amsmath}	% Advanced maths commands
\newcommand{\tfifty}{$t_{50}$ }

\newcommand{\msun}{{\rm M}_\odot}

\title[Stellar Metallicity Gradients in Dwarf Galaxies]{A Relationship Between Stellar Metallicity Gradients and Galaxy Age in Dwarf Galaxies}

\author[F. J. Mercado et al.]
{\parbox{17.5cm}{
Francisco J. Mercado$^{1}$\thanks{E-mail: mercadf1@uci.edu}, James S. Bullock$^{1}$, Michael Boylan-Kolchin$^{2}$, Jorge Moreno$^{3}$, Andrew Wetzel$^{4}$, Kareem El-Badry$^{5}$, Andrew S. Graus$^{2}$, Alex Fitts$^{2}$, Philip F. Hopkins$^{6}$, Claude-Andr{\'e} Faucher-Gigu{\`e}re$^{7}$, Alexander B. Gurvich$^{7}$ }\vspace{0.3cm}\\
$^{1}$Center for Cosmology, Department of Physics and Astronomy,4129 Reines Hall, University of California Irvine, CA 92697, USA \\
$^{2}$Department of Astronomy, The University of Texas at Austin, 2515 Speedway Stop C1400, Austin, TX 78712, USA\\
$^{3}$Department of Physics and Astronomy, Pomona College, Claremont, CA 91711, USA\\
$^{4}$Department of Physics, University of California, Davis, CA 95616, USA\\
$^{5}$Department of Astronomy and Theoretical Astrophysics Center, University of California Berkeley, Berkeley, CA 94720, USA\\
$^{6}$TAPIR, Mailcode 350-17, California Institute of Technology, Pasadena, CA 91125, USA\\
$^{7}$Department of Physics and Astronomy and CIERA, Northwestern University, 2145 Sheridan Road, Evanston, IL 60208, USA\\
}

\date{Accepted XXX. Received YYY; in original form ZZZ}

\pubyear{2020}

\begin{document}
\label{firstpage}
\pagerange{\pageref{firstpage}--\pageref{lastpage}}
\maketitle

%%%%%%%%%%%%%%%%%%%%%%%%%%%%%%%%%%%%%%%%%%%%%%%%%%%%%%%%%%%%%%%%%%%%%
%%%%%%%%%%%%%%%%%%%%%%%%%%% ABSTRACT %%%%%%%%%%%%%%%%%%%%%%%%%%%%%%%%
%%%%%%%%%%%%%%%%%%%%%%%%%%%%%%%%%%%%%%%%%%%%%%%%%%%%%%%%%%%%%%%%%%%%%

\begin{abstract}
We explore the origin of stellar metallicity gradients in simulated and observed dwarf galaxies. We use FIRE-2 cosmological baryonic zoom-in simulations of 26 isolated galaxies as well as existing observational data for 10 Local Group dwarf galaxies. Our simulated galaxies have stellar masses between $10^{5.5}$ and $10^{8.6} \msun$. Whilst gas-phase metallicty gradients are generally weak in our simulated galaxies, we find that stellar metallicity gradients are common, with central regions tending to be more metal-rich than the outer parts. The strength of the gradient is correlated with galaxy-wide median stellar age, such that galaxies with younger stellar populations have flatter gradients. Stellar metallicty gradients are set by two competing processes: (1) the steady ``puffing'' of old, metal-poor stars by feedback-driven potential fluctuations, and (2) the accretion of extended, metal-rich gas at late times, which fuels late-time metal-rich  star formation. If recent star formation dominates, then extended, metal-rich star formation washes out pre-existing gradients from  the ``puffing'' process. We use published results from ten Local Group dwarf galaxies to show that a similar relationship between age and stellar metallicity-gradient strength exists among real dwarfs.  This suggests that observed stellar metallicity gradients may be driven largely by the baryon/feedback cycle rather than by external environmental effects.

\end{abstract}


\begin{figure*}
	\includegraphics[width=1.0\textwidth, trim = 0 0 0 0]{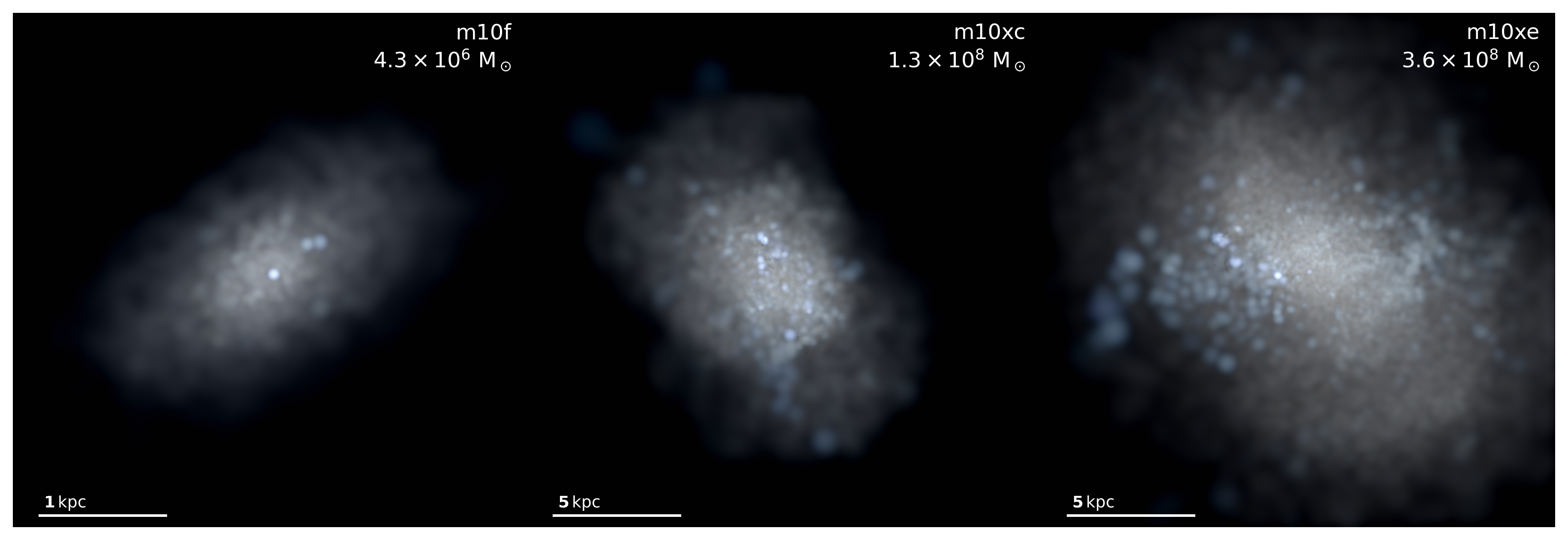}
    
	\centering
	\caption[examples]{--- \textit{\textbf{Mock Hubble Space Telescope Images.}} Mock u/g/i composite images of three galaxies within our sample (m10f, m10xc, and m10xe from left to right). We choose these galaxies to represent the full range in galaxy sizes and stellar masses (top right) within our sample. The average projected half-light radii of these galaxies are $R_{1/2} = 0.52$ kpc, 2.25 kpc, and 2.93 kpc, from left to right, respectively.}
	\label{fig:mock_hubble}
\end{figure*}

%%%%%%%%%%%%%%%%%%%%%%%%%%%%%%%%%%%%%%%%%%%%%%%%%%%%%%%%%%%%%%%%%%%%%%%%%%%%%%%%%
%%%%%%%%%%%%%%%%%%%%%%%%%%%%%%%% INTRODUCTION %%%%%%%%%%%%%%%%%%%%%%%%%%%%%%%%%%%
%%%%%%%%%%%%%%%%%%%%%%%%%%%%%%%%%%%%%%%%%%%%%%%%%%%%%%%%%%%%%%%%%%%%%%%%%%%%%%%%%

\section{Introduction}
Dwarf galaxies are critical laboratories for understanding galaxy assembly over cosmic time. Within the Local Group, dwarf galaxies can be split into two categories. Dwarf Irregulars (dIrrs) are gas-rich, generally star forming systems that are commonly found in isolation \citep{Mateo1998, McConnachie12, Simon19}. Dwarf Spheroidals (dSphs) are gas-poor systems that exhibit little to no recent star formation and are, in general, satellites orbiting larger galaxies \citep{Mateo1998, Grebel1999, Wilkinson2004, Munoz2005,Munoz2006b,Walker2006, Walker2007, Koch2007a, Mateo2008}. Whilst dIrr and dSph galaxies are found in  different environments and have some distinct characteristics, they fall on the same relationship between stellar mass and average stellar metallicity \citep{Kirby13} and follow similar scaling relations \citep{Tolstoy09}. Thus, the extent to which a dwarf galaxy's current appearance and morphology are affected by external environment, as opposed to internal processes (such as supernova feedback and gas cooling), remains an active area of research. 

The heavy element content of dwarf galaxies provides a means to explore feedback processes and metal retention in systems with relatively shallow potential wells \citep{DekelSilk86,Mateo1998}. In particular, dwarf galaxies often exhibit radial variations in stellar metallicity, with more metal-poor populations at large galactocentric radii. Such gradients provide an opportunity to explore relationships between feedback and dynamical evolution in shaping small galaxies \citep{Koleva11}. The strength of the gradient can vary significantly amongst galaxies with approximately the same stellar mass  \citep{Saviane01, Harbeck01, Tolstoy04, Battaglia06, Battaglia11, Kirby10, Koleva11,Vargas14, Ho15, Kacharov17}. If we can identify systematic trends underlying these differences, this may provide insight into their origin. 

Recent studies of stellar metallicity gradients in Local Group dwarf galaxies point to a slight dichotomy in the strengths of the galaxy gradients, such that (older, more dispersion-supported) dSphs commonly exhibit stronger gradients than dIrrs \citep{Leaman13, Kacharov17}. A potential reason for this difference is proposed by \citet{Schroyen11}, who use idealised dwarf galaxy simulations to show that systems initialised with higher angular momentum tend to have weaker gradients. Specifically, the younger stellar populations formed from enriched gas with high angular momentum tend to be spatially smoother (i.e. less radially clustered), flattening those galaxies' stellar metallicity gradients.  Other internal mechanisms that can play a role in the existence of gradients include feedback-induced redistribution of material in the ISM \citep{De94} or the perturbation of stellar orbits \citep{Read2005,Pontzen2012,ElBadry16} due to potential fluctuations. A third process potentially responsible for affecting a galaxy's gradient is ram-pressure stripping of an infalling satellite (important for LG dSphs). In this scenario, the radial extent of star formation in satellite galaxies is reduced, enriching the central regions as a result \citep{Mayer01, Mayer07}.

In this work we analyse 26 cosmological zoom simulations of isolated dwarf galaxies with stellar masses from  $10^{5.5}-10^{8.6} \, \msun$ from \citet{Graus19} and \citet{Fitts17}, run using \texttt{GIZMO} \citep{Hopkins15} with the FIRE-2 feedback implementation \citep{Hopkins18}. We show that a range of stellar metallicity gradients arises naturally in these isolated systems and that the strength of the gradient correlates with overall galaxy age, such that galaxies that form a larger fraction of their stars at late cosmic times have flatter gradients. Gradients arise from the steady redistribution of old, metal-poor stars from central, feedback-driven potential fluctuations and can be weakened when late-time star formation is sourced by recycled (enriched) gas, which is deposited at larger radii. These trends appear consistent with what has been observed in Local Group galaxies, without appealing to environmental processes. 

This paper is organised as follows: In Section \ref{sec:sims} we discuss our simulations. In Section \ref{sec:results} we present stellar metallicity gradient measurements in our simulated sample and show how gradient strengths increase with galactic ages. Note that all instances of galaxy age in this paper refer to the lookback age of the galaxy (i.e. not cosmic age). Finally, in Sections \ref{sec:origin} \& \ref{sec:obs} we discuss the mechanisms that form and set a galaxy's gradient strength and present a qualitative comparison between the gradient-strength-galaxy-age relations for our simulated sample and 10 Local Group dwarf galaxies.

%%%%%%%%%%%%%%%%%%%%%%%%%%%%%%%%%%%%%%%%%%%%%%%%%%%%%%%%%%%%%%%%%%%%%%%%%%%%%%%%%
%%%%%%%%%%%%%%%%%%%%%%%%%%%%%% SIMULATIONS %%%%%%%%%%%%%%%%%%%%%%%%%%%%%%%%%%%
%%%%%%%%%%%%%%%%%%%%%%%%%%%%%%%%%%%%%%%%%%%%%%%%%%%%%%%%%%%%%%%%%%%%%%%%%%%%%%%%%

\section{Simulations}

\begin{figure*}
	\includegraphics[width=1.0\textwidth, trim = 0 0 0 0]{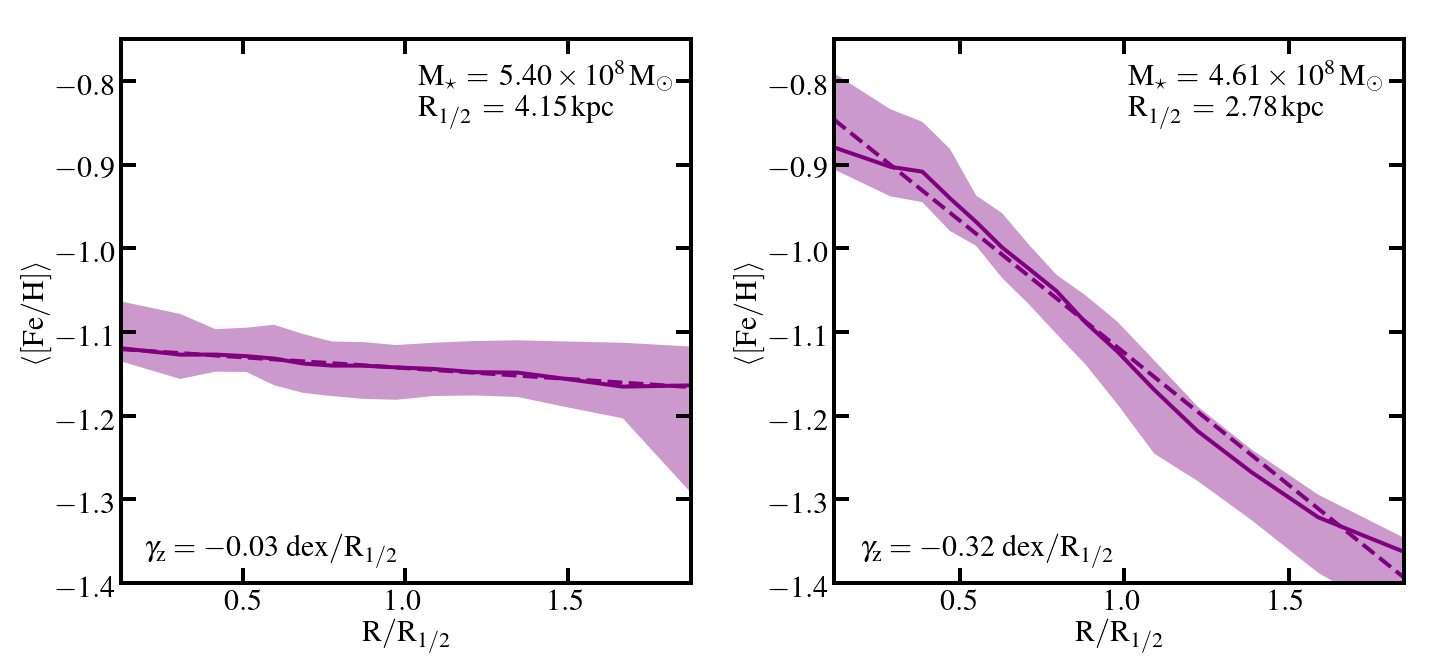}
    
	\centering
	\caption[examples]{--- \textit{\textbf{Two example gradients.}} Median stellar iron abundance as a function of projected 2D radius $R$ for two example galaxies, one with a weak stellar metallicity gradient (m10xh; left) and one with a strong gradient (m10xg; right). Each panel shows information measured over 100 random viewing angles. The solid line represents the median metallicity at a given radius over all projections, whilst the shaded region spans the full range of median metallicities measured at a given radius over all projections. The dashed line represents the least squares fit to the solid line and its slope is taken to be the galaxy's stellar median metallicity gradient ($\gamma_{z}$; value shown in the lower left). These galaxies were chosen to represent the range of gradient strengths within our sample.}
	\label{fig:ex_profs}
\end{figure*}

Our simulations employ the gravity+hydrodynamic code \texttt{GIZMO} \citep{Hopkins15} with the Meshless Finite Mass (MFM) hydrodynamic solver and FIRE-2 (Feedback In Realistic Environments\footnote{The FIRE project website: \href{https://fire.northwestern.edu/}{https://fire.northwestern.edu/} }) feedback implementation \citep{Hopkins18}. MFM offers many advantages over classic SPH, including the ability to capture mixing instabilities and resolve sharp shocks, to accurately evolve sub-sonic flow with minimal numerical viscosity, all while conserving energy and angular momentum to a high degree of accuracy (Hopkins 2015).

The simulations include gas cooling due to molecular transitions and metal-line fine structure transitions at low temperatures whilst cooling at temperatures of $\geq 10^4 K$  is due to primordial and metal line cooling. We adopt a redshift-dependent, ionising UV background model from \citet{FG2009}. Star formation occurs for self-shielding, molecular gas that is above a threshold density of $n_{\rm crit} \geq 1000\, \rm cm^{-3}$, self-gravitating, and Jeans unstable \citep[see][]{Hopkins18} for details). Once a star particle is formed, it is treated as a single stellar population with a Kroupa initial mass function \citep{Kroupa02} and a mass and metallicity inherited from its progenitor gas particle. The total metallicity along with eleven chemical species (H, He, C, N, O, Ne, Mg, Si, S, Ca, Fe) are tracked for each gas and star particle -- of which Fe and H are the most relevant to this work. Feedback mechanisms include supernovae (Ia \& II), stellar mass-loss due to fast and slow winds (from OB-stars and AGB-stars, respectively), photo-ionisation/electric heating, and radiation pressure. The ensuing feedback quantities are calculated from stellar population models  \citep[][STARBURST99]{Leitherer99}. Sub-grid metal diffusion is implemented to account for the turbulent eddies between gas particles in a turbulent ISM. Without such an implementation the metals assigned to a given gas particle would remain locked in with that particle throughout time, not allowing for proper enrichment. Whilst this does not affect the mean metallicity of a galaxy, this implementation has been shown to produce more realistic metallicity distribution functions in simulated low mass galaxies \citep{Escala18}. We note that whilst previous FIRE work suggests that galaxies simulated with FIRE-2 physics tend to exhibit average stellar metallicities lower than that seen in real galaxies, the slope of the Mass-Metallicity Relation of FIRE-2 simulated galaxies seems to match that of the observed galaxies \citep{Escala18, Wheeler2019}. Furthermore, this work focuses on the \textit{slope} of stellar metallicity profiles in galaxies rather than their normalization and we therefore believe that the underprediction of low-mass metallicities by FIRE-2 is not an issue for this specific study.

\begin{figure*}
	\includegraphics[width=1.0\textwidth, height=0.31
	\textheight,, trim = 0 0 0 0]{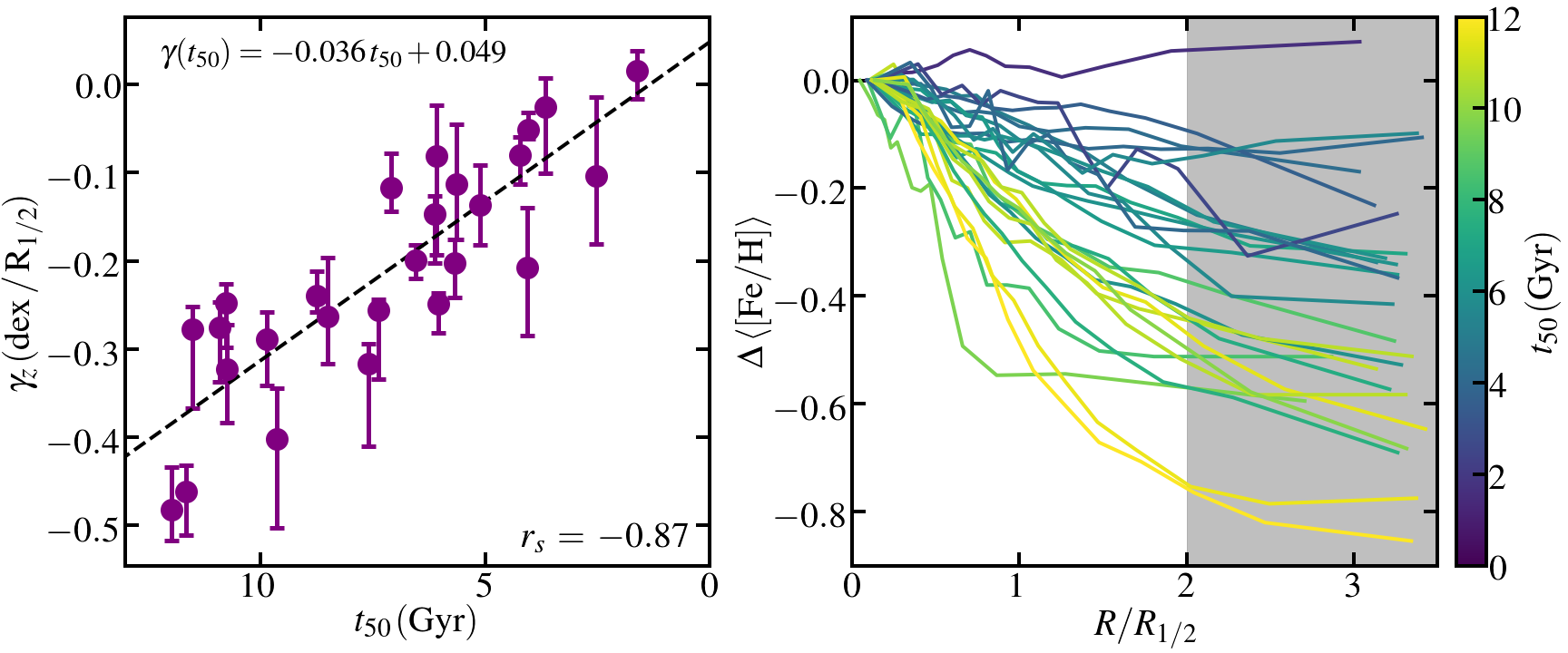}
	\centering
	\caption[gradients and ages]{--- \textit{\textbf{The simulated gradient-strength-galaxy-age relationship.}} A correlation between galactic age and stellar metallicity gradient strength in simulated galaxies. {\em Left:} Stellar metallicity gradient strength for each galaxy versus median age of stars in that galaxy ($t_{50}$). The error bars correspond to the range of metallicity gradient strengths measured over 100 random viewing angles. The black dashed line represents the least squares fit to the relationship (presented at the top right). The slope and y-intercept of the fit are $-0.036 \pm 0.005$ and $0.049 \pm 0.035$, respectively. At the bottom right, we provide the spearman coefficient ($r_{s} = -0.87$) calculated for this relationship. {\em Right:} Median iron abundance (measured over all projections) for each galaxy plotted as a function of projected radius in units of the 2D half-mass radius. The profiles are normalised by the metallicity at the centre of each galaxy and are coloured corresponding to their $t_{50}$ values, with yellow representing earlier star formation times (older galaxies) and purple representing more recent star formation times (younger galaxies). We use the profile data within $2R_{1/2}$ to calculate the gradient strengths and thus exclude the part of the profiles that extend into the shaded region. In both panels older galaxies tend to have steeper gradients.}
	\label{fig:all_grad}
\end{figure*}

\citet{Graus19} and in \citet{Fitts17} were the first to present the galaxies analysed in this work. 
The \citet{Graus19} galaxies (m10xa-i) consist of galaxies with $M_{\rm vir} \simeq$ $10^{10}$ $\msun$ -- $10^{11}$ $\msun$ and stellar masses between $10^{7}$ $\msun$ and $10^{9}$ $\msun$. These simulations were run with a dark matter particle mass of $m_{\rm dm}$ = 20000 $\msun$ and an initial gas particle mass of $m_{\rm g}$ = 4000 $\msun$. The \citet{Fitts17} galaxies (m10b-m) consist of lower mass galaxies with $M_{\rm vir} \simeq$ $10^{10}$ $\msun$ and stellar masses between $10^{5}$ $\msun$ and $10^{7}$ $\msun$. These simulations were run with dark matter particle masses of $m_{\rm dm}$ = 2500 $\msun$ and gas particles with an initial particle mass of $m_{\rm g}$ =  500 $\msun$. Finally, \citet{Graus19} adopts the cosmological parameters: $H_0$ = 70.2 km s$^{-1}$ Mpc$^{-1}$, $\Omega_{\rm m}$ = 0.272, $\Omega_{\rm b}$ = 0.0455, $\Omega_{\Lambda}$ = 0.728, whilst the \citet{Fitts17} simulations were run with a slightly different set of cosmological parameters: $H_0$ = 71.0 km s$^{-1}$ Mpc$^{-1}$, $\Omega_{\rm m}$ = 0.266, $\Omega_{\rm b}$ = 0.044, $\Omega_{\Lambda}$ = 0.734.

We provide mock \textit{Hubble Space Telescope} images of three of our simulated galaxies in Figure \ref{fig:mock_hubble} (m10f, m10c, m10xe). These examples span the range of sizes and masses of the galaxies in our sample. In what follows we explore two measures of metallicty gradients: one in 2D, $\gamma_{z}$, and one in 3D, $\widetilde{\gamma}_{z}$.  The projected gradients are used to connect with observations while the 3D gradients are used to infer physical meaning.  Given the range of galaxy sizes, we measure gradients in units of each galaxy's half-mass radius in order to compare galaxies across our full sample.  It is clear from Figure \ref{fig:mock_hubble} that these galaxies are aspherical.  However, for the sake of simplicity, below we use spherically-averaged 3D gradients and circularly-averaged 2D gradients measured over many viewing angles. We also restrict our quantitative measures to within two times the half-light radii of galaxies, regions where the systems are somewhat more spherical, as can be inferred from Figure \ref{fig:mock_hubble}.

\label{sec:sims} % used for referring to this section from elsewhere

%%%%%%%%%%%%%%%%%%%%%%%%%%%%%%%%%%%%%%%%%%%%%%%%%%%%%%%%%%%%%%%%%%%%%%%%%%%%%%%%%%%%%%%%%%%%%%%%%%%%%%%%%%%%%%%%
%%%%%%%%%%%%%%%%%%%%%%%%%%%%%%%%%% RESULTS: AGE AND METALLICITY GRADIENTS %%%%%%%%%%%%%%%%%%%%%%%%%%%%%%%%%%%%%%
%%%%%%%%%%%%%%%%%%%%%%%%%%%%%%%%%%%%%%%%%%%%%%%%%%%%%%%%%%%%%%%%%%%%%%%%%%%%%%%%%%%%%%%%%%%%%%%%%%%%%%%%%%%%%%%%

\section{Results: Age and Stellar Metallicity Gradients}

\begin{figure*}
	\includegraphics[width=1.0\textwidth,height=0.74
	\textheight, trim = 0 0 0 0]{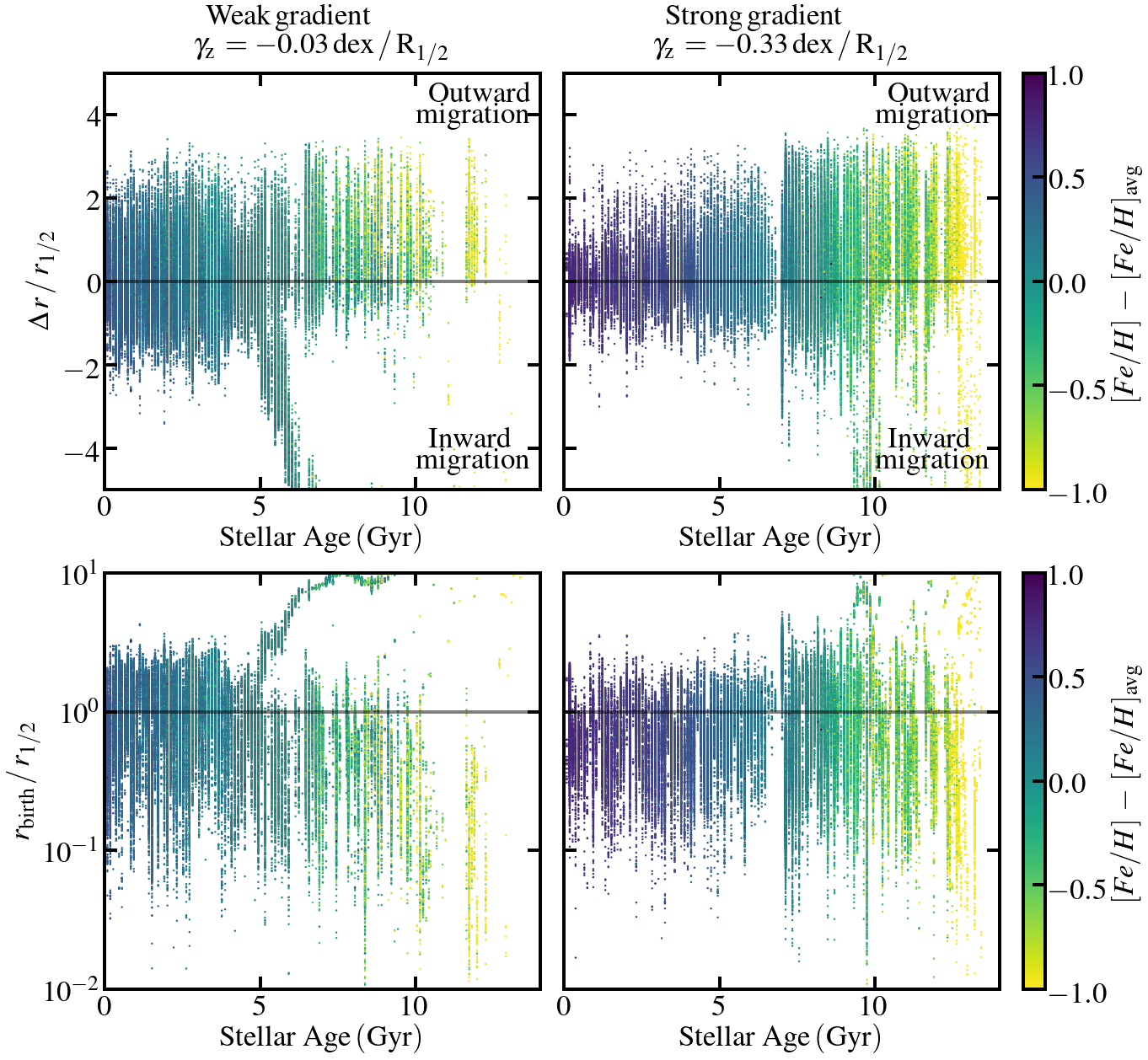}
	\hspace{.1in}
	\centering
	\caption[Migration]{--- \textit{\textbf{Stellar radial migration and birth radius versus stellar age.}} Radial migration of stellar populations with age for the same weak-gradient (left) and strong-gradient (right) examples as in Figure \ref{fig:ex_profs}. We colour star particles by their iron-to-hydrogen abundance ratio normalised with respect to the galaxy average. {\em Top panels:} net change in stellar particle radial position since its birth ($\Delta r = r_{\rm now} - r_{\rm birth}$) in units of $r_{1/2}$ at $z=0$, versus age. The horizontal black line corresponds to no net radial migration. Points above the line have migrated outward. In general, {\em the older (more metal-poor) star particles experience more outward migration than their younger (more metal-rich) counterparts.} This is true for both the weak-gradient and strong-gradient examples. {\em  Bottom panels:} the birth radius of each star particle in units of the galaxy's $r_{1/2}$ versus age. Whilst both examples show a trend between radial migration and stellar age, the weak gradient case experiences more radially extended late-time star formation, which acts to flatten the gradient. Note that the features at $\sim$6 Gyr (left) and $\sim$10 Gyr (right) that trail downward in the top panels and upward in the bottom panels are mergers. We find that mergers play at best a secondary role in shaping metallicity gradients.}
	\label{fig:migration}
\end{figure*}

Figure \ref{fig:ex_profs} presents two representative stellar metallicity gradients from the simulated sample: m10xh (left; weak gradient) and m10xg (right; strong gradient). Whilst the galaxies have comparable stellar masses and similar mean metallicities, they have markedly different radial variations in stellar [Fe/H]. Each panel shows the iron abundance profile measured in annular bins of 2D projected radius $R$. The solid lines show the median metallicity measured at a given radius over 100 random viewing angles. The shaded region represents the full range of median stellar metallicity at a given radius over all viewing angles. The dashed line shows a least-squares fit to the solid line and its slope is taken to be the galaxy's metallicity gradient, $\gamma_{z}$. The gradients are normalised by the mean 2D half-mass radius over all projections and thus have units of dex per $R_{1/2}$. We examine many projections of our simulated galaxies to account for the fact that the galaxies are not spherical, as evidenced by Figure \ref{fig:mock_hubble}. The plot shows a clear radial variation in the metallicity of the galaxy such that the inner (outer) regions of the galaxy are primarily populated by stars of higher (lower) metallicity.

We list the projected stellar metallicity gradients, along with other properties, for all galaxies in our sample in Table \ref{table:galaxy_stats}. The majority of these galaxies have clear negative gradients with more metal-rich stars residing in the centers of the galaxies and more metal-poor stars occupying the outer regions. \citet{Graus19} finds similar (also negative) gradients in the ages of the stellar populations of the galaxies in this sample. The significant variation in $\gamma_{z}$ from galaxy to galaxy motivates us to seek correlations in this variable with other galaxy properties. As we show in the Appendix, we find that $\gamma_z$ is largely independent of halo mass, galaxy mass, and galaxy size. However, we find a strong correlation with galaxy age (as well as age gradient). Finally, we find no clear correlation between a galaxy's gradient strength and the its $v/\sigma$ value. This is in contradiction with results from \citet{Schroyen11} that suggest that rotaion-supported galaxies tend to exhibit stronger stellar metallicity gradients -- though it is important to note that the simulated galaxies in our sample have, overall, lower values of $v/\sigma$ than the galaxies considered in \citet{Schroyen11}.

Figure \ref{fig:all_grad} depicts a relationship between metallicity gradient-strength and galactic age in two ways. On the left the gradients, $\gamma_{z}$, are depicted as a function of $t_{50}$  -- the fiftieth percentile age of the stars in a given galaxy, i.e. the median age. Our sample follows the following relationship:
\begin{equation}
    \gamma(t_{50}) = -0.036 t_{50} + 0.049,
\end{equation}
with the slope and y-intercept having errors of $\pm 0.005$ and $\pm 0.035$, respectively. We provide the calculated spearman coefficient, $r_{s}$, as a measure of how well the two variables are correlated. On the right we show the mean stellar iron abundance as a function of projected radius (as in Figure \ref{fig:ex_profs}) for every simulated galaxy in our sample, colour-coded by the galaxy's median stellar age (i.e. lookback age). The profiles are normalised by the metallicity at the centre of each galaxy. Yellow and purple represent older and younger stellar populations, respectively. We use the profile data within $2R_{1/2}$ to calculate the stellar metallicity gradients. Thus, the parts of the profiles that extend into the grey shaded region are excluded. In general, \textit{the galaxies that form their stars earlier tend to have stronger stellar metillicity gradients}. We explore correlations with other measures of galaxy age ($90$th percentile age, $25$th percentile age, etc.) and find the correlation is strongest when using the stellar component's median age. 

\citet{Graus19} find a similar relationship for age gradients as a function of galaxy age. We find that metallicity gradients appear to correlate even more tightly with galaxy age than do age gradients. This suggests that stellar metallicities can serve as a better "internal dynamical clock" for galaxy gradients than do absolute age gradients.

\label{sec:results}
%%%%%%%%%%%%%%%%%%%%%%%%%%%%%%%%%%%%%%%%%%%%%%%%%%%%%%%%%%%%%%%%%%%%%%%%%%%%%%%%%%%%%%%%%%%%%%%%%%%%%%
%%%%%%%%%%%%%%%%%%%%%%%%%%%%%%%% ORIGIN OF METALLICITY GRADIENTS %%%%%%%%%%%%%%%%%%%%%%%%%%%%%%%%%%%%%
%%%%%%%%%%%%%%%%%%%%%%%%%%%%%%%%%%%%%%%%%%%%%%%%%%%%%%%%%%%%%%%%%%%%%%%%%%%%%%%%%%%%%%%%%%%%%%%%%%%%%%

\section{Origin of Stellar Metallicity Gradients}

\begin{figure}
	\includegraphics[width=\columnwidth, height=0.32
	\textheight,, trim = 0 0 0 0]{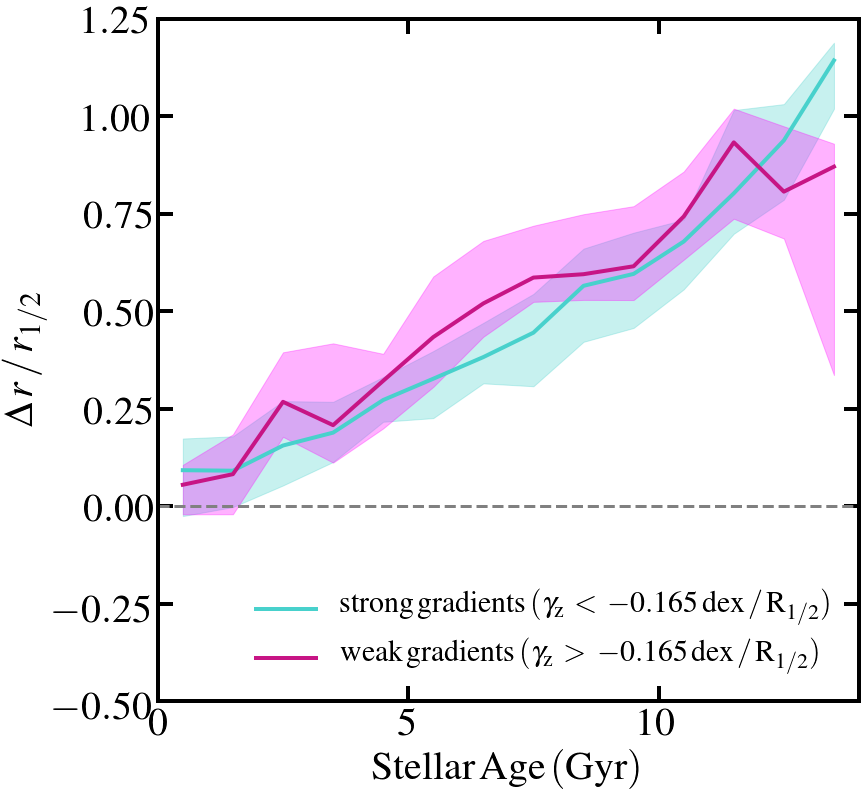}
	\centering
	\caption[migration all]{--- \textit{\textbf{Stellar population migration versus stellar age.} }The migration of stars, $\Delta \, r$, in units of the $z=0$ 3D half-mass radius as a function of stellar age for the simulated galaxies in the \citet{Graus19} sample. We divided the sample into galaxies with strong gradients ($\gamma_{z} <-0.165\, dex/R_{1/2}$) and galaxies with weak gradients ($\gamma_{z} >-0.165\, dex/R_{1/2}$), where $-0.165\, dex/R_{1/2}$ is the median value of $\gamma_{z}$. The shaded regions represent the 68th percentile spread of the data. It is clear that older stars have, in general, migrated farther out than their younger counterparts in the entire simulation sample, regardless of the galaxy's gradient strength.}
	\label{fig:migration_all}
\end{figure}

Dynamical heating associated with stellar feedback appears to play an important role in creating the gradients we see in our simulated galaxies. Note that star formation in FIRE dwarf galaxies is particularly time-variable throughout cosmic time \citep{Sparre2017, FG2018, FV2020}. \citet{ElBadry16} use a separate suite of FIRE simulations to explore the role of stellar feedback on radial migration of stars. They find that older stars generally experience more outward migration than their younger counterparts. Repeated bursts of star formation throughout a galaxy's lifetime create gravitational interactions that systematically drive stars towards more extended orbits \citep[see Figures 4 \& 5 of][]{ElBadry16}. This results in a scenario in which stars that live through more ``puffing cycles'' will have migrated more, on average, than younger stars that live through fewer cycles. We note that the effects of bursty stellar feedback on stellar distributions is analogous to the effects of stellar feedback on the dark matter distribution in halo centres \citep{Read2005,Pontzen2012, Onorbe2015, Chan2015}.

The top panels of Figure \ref{fig:migration} illustrate the process of stellar migration. We show the change in each star particle's radial position since its birth ($\Delta r = r_{\rm now} - r_{\rm birth}$), normalised by the galaxy 3D half-mass radius today, $r_{1/2}$, as a function of that star's age. Positive values of $\Delta r$ indicate that the star has moved outward. The left and right panels show the same information for the same weak-gradient example (m10xh) and strong-gradient example (m10xg) respectively (as in Figure \ref{fig:ex_profs}). Points are colour-coded by the iron-to-hydrogen abundance ratio assigned to that particle, normalised by the average [Fe/H] of the galaxy. In both the weak-gradient and strong-gradient galaxies, the oldest, most metal-poor stars tend to migrate outward, more than their younger, more metal-rich counterparts. The difference between the two galaxies is in the younger, more metal-rich population. The galaxy with the weaker gradient (left) has a significant amount of late-time, metal-rich star formation occurring at large radii, whilst the strong-gradient case (right) has all the new star formation confined to a small radii. Note that there are features at $\sim$6 Gyr (left panel) and $\sim$10 Gyr (right panel) indicative of mergers (large negative values of $\Delta r$). Whilst there is a possibility for a galaxy's gradient to be altered by merger processes -- through the deposition of old, metal-poor stars onto a galaxy's outskirts -- this does not seem to be the case for the galaxies in our sample. For the galaxies in our sample, the fraction of mass in stars that is acquired through mergers is negligible and distributed throughout the whole galaxy and thus will play, at best, a minor role in changing the galaxy's stellar metallicity gradient \citep{Fitts18, Graus19}. This is in line with results from recent investigations that suggest that morphological disturbances in dwarf galaxies are seldom driven by mergers \citep{Martin2020}.

\begin{figure}
	\includegraphics[width=\columnwidth, height=0.32
	\textheight,, trim = 0 0 0 0]{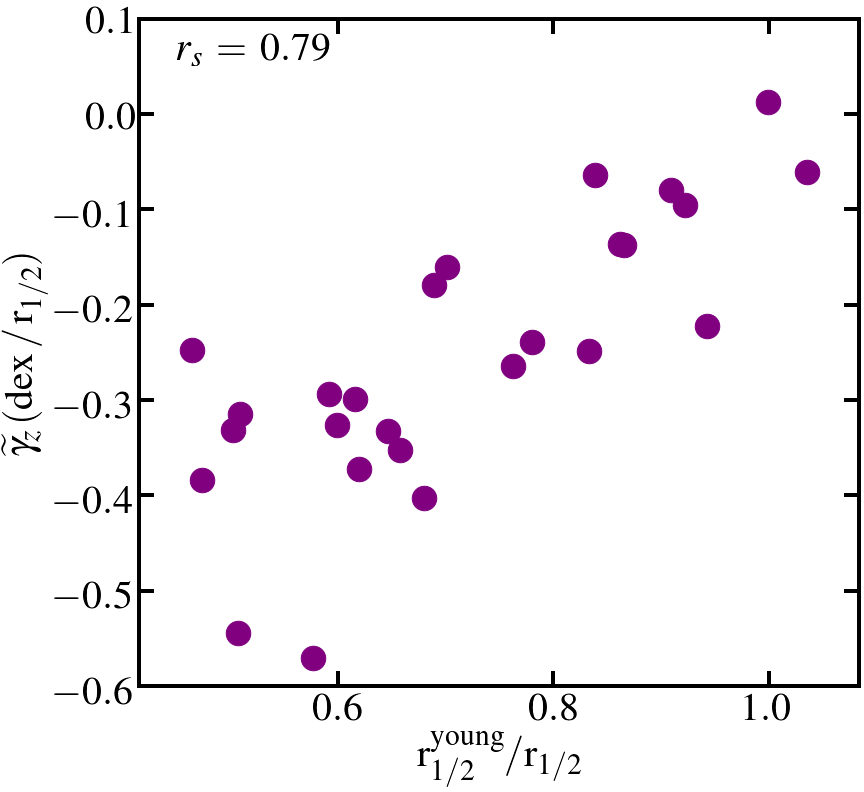}
	\centering
	\caption[young stars]{--- \textit{\textbf{Gradient strength versus young stellar population size.}} The 3D stellar metallicity gradient strength as a function of median radial position (in units of $r_{1/2}$) of the star particles younger than 4 Gyrs old within $4 r_{1/2}$ of their respective galaxy. There is a correlation ($r_{s} = 0.86$) in which galaxies with stronger gradients have a young stellar population that is more centrally concentrated than the young stellar populations of galaxies with weaker gradients.}
	\label{fig:young_stars}
\end{figure}

\begin{figure*}
	\includegraphics[width=1.0\textwidth, height=0.33
	\textheight,, trim = 0 0 0 0]{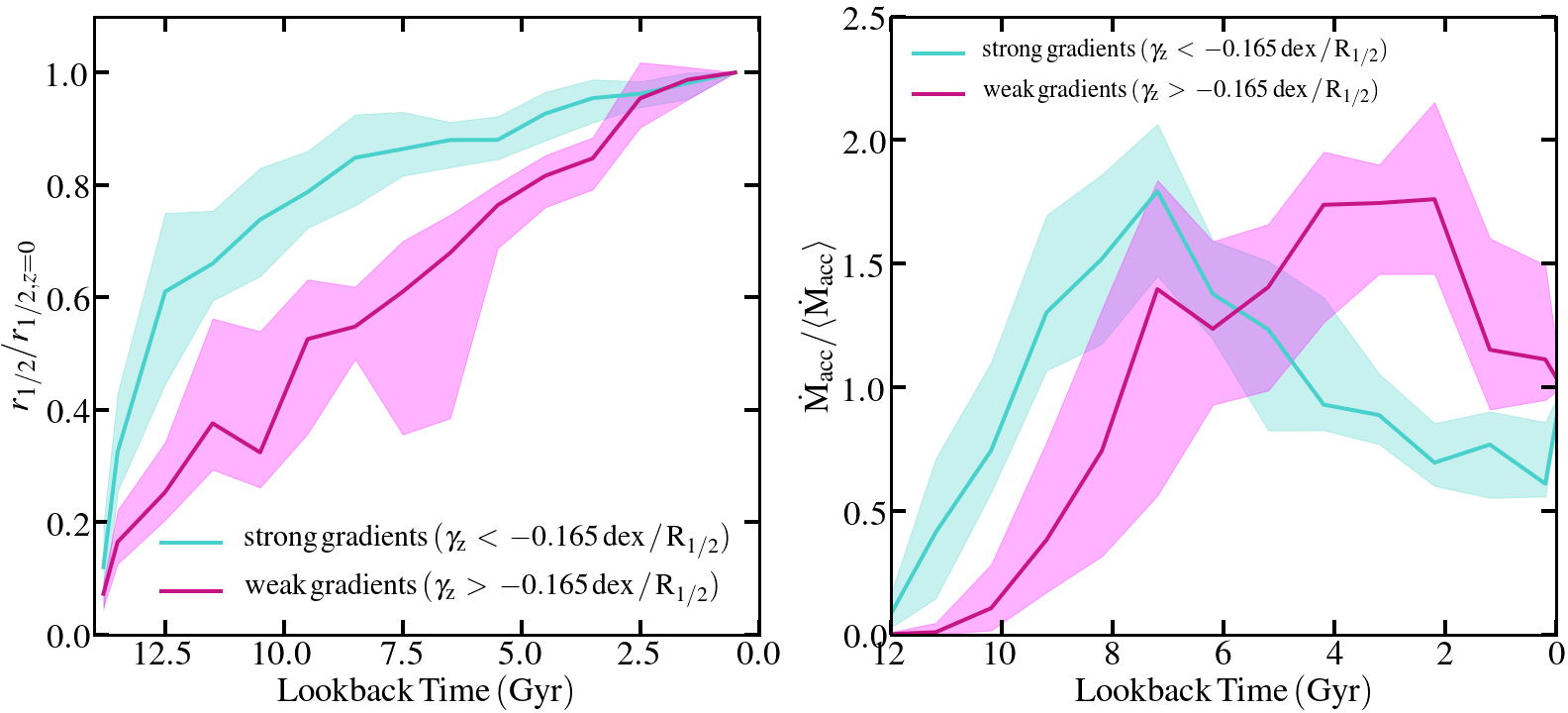}
	\centering
	\caption[gas and size evo]{--- \textit{\textbf{The size evolution and gas accretion history of our simulated galaxies.}} Two factors that likely contribute to setting the strength of a galaxy's gradient. {\em Left:} The half-mass radius value at a given time normalised by the half-mass radius today as a function of lookback time. {\em Right:} The gas mass accretion rate at a given time, normalised by the time averaged gas mass accretion rate in a given galaxy, versus lookback time. The data are split into two categories -- galaxies with strong gradients (cyan) and galaxies with weak gradients (magenta). The shaded regions represent the 68th percentile spread of the simulated data. Galaxies with strong gradients tend to set their size and experience more gas accretion earlier on than do galaxies with weak gradients.}
	\label{fig:gas_all}
\end{figure*}

The bottom panels of Figure \ref{fig:migration} provide a complementary picture. The vertical axis displays the ratio of a star's birth radius to the current half-mass radius versus stellar age and colour-coded by metallicity. One important takeaway from these panels is that the particles in the same age bin indicate no clear radial metallicity gradient. This implies that there was no radial variation in the metallicity of the cold, dense, and self-gravitating gas making stars at any given time. This rules out the possibility of a pre-existent gas-phase gradient ultimately driving the global stellar metallicity gradient we witness at late times. This is consistent with results from \citet{Escala18} which show that FIRE-2 dwarf galaxies have ISMs that are well mixed at any given time. Second, in the weaker gradient case (left), late-time star formation is occurring at much larger radii than early star formation. This allows the formation of metal-rich stars at large radii, which mitigates the effect of migration of metal-poor stars from feedback puffing. The strong-gradient galaxy (right) lacks late-time extended star formation, which preserves the gradient established by migration of old stars.

Figure \ref{fig:migration_all} shows the change in each star particle's radial position since its birth ($\Delta r = r_{\rm now} - r_{\rm birth}$) for all galaxies in the \citet{Graus19} sample, normalised by the 3D half-mass radius today, $r_{1/2}$, as a function of that star's age\footnote{Only the \citet{Graus19} galaxies were used in this figure (as well as Figure \ref{fig:gas_all}) because only the $z=0$ snapshots of the \citet{Fitts17} sample were saved. We do not expect there to be much difference in the \citet{Fitts17} sample as they span the same range of metallicity gradients and lie on the same gradient-strength-galaxy-age relationship.}. We divide the galaxies in two groups: strong gradients ($\gamma_{z} < -0.165\, dex/R_{1/2}$, cyan) and weak gradients ($\gamma_{z} > -0.165\, dex/R_{1/2}$, magenta) where $-0.165 \, dex/R_{1/2}$ is the median value of $\gamma_{z}$ for the sample. The solid lines show the median value of $\Delta\,r/r_{1/2}$ in a given stellar age bin for the galaxies in a given group. The shaded regions represent the 68th percentile spread about the median. It is clear that regardless of a galaxy's gradient strength at $z=0$, older, more metal-poor star particles migrate farther out than their younger, more metal-rich counterparts. This suggests that \textit{the difference between weak and strong gradient galaxies must be connected to how the young, metal-rich stars are distributed, rather than differences in the migration of old, metal-poor stars.}

In Figure \ref{fig:young_stars} we illustrate the systematic role of late-time star formation in flattening gradients. We show the 3D stellar metallicity gradients for our full sample ($\widetilde{\gamma}_{z}$) versus the ratio of $r_{1/2}^{\rm young}$ to $r_{1/2}$, where $r_{1/2}^{\rm young}$ is defined by the median 3D radial position of stars born within the last 4 Gyrs. Here we normalize the gradients by the galaxy stellar 3D half mass radius (as opposed to the 2D radius-normalized gradients in Figure \ref{fig:ex_profs}) in order to better-examine the internal dynamics of the system. We include every star within 4$r_{1/2}$ to capture the region in which most star formation is happening. There is a clear relationship ($r_{s} = 0.86$) between a galaxy's gradient strength and the radius of recent star formation. \textit{Galaxies with weaker gradients have more extended young stellar populations, whilst galaxies with stronger gradients have recent star formation confined to the centre}. This suggests that metallicity gradients are flattened in galaxies that undergo radially extended, late-time star formation.

Figure \ref{fig:gas_all} demonstrates that galaxies with strong gradients (cyan) tend to have their sizes set earlier (left) and undergo more gas accretion early (right) than do galaxies with weak gradients (magenta). In the left panel we are plotting the stellar half-mass radius of the main progenitor of each galaxy, normalised by the half mass radius today, as a function of lookback time. As in Figure \ref{fig:migration_all}, we divide galaxies into strong and weak gradient samples based on their $z=0$ gradients. The solid lines show the median relation for galaxies in each sample. The shaded regions represent the 68th percentile spread over the sample.

The solid lines in the right-hand panel of Figure \ref{fig:gas_all} show the median gas mass accretion rate
for all galaxies normalised by the galaxy's time-averaged gas mass accretion rate. The shaded regions represent the 68th percentile spread over the sample. 
We compute the gas mass accretion rate within $0.1R_{\rm vir}$ using the following formula \citep{FG2011,Muratov2015}:
\begin{equation}
    \dot{\rm M}_{\rm acc} = \frac{\partial \rm M}{\partial t} =  \frac{\sum v_{\rm rad} \, m_{\rm p}}{0.1R_{\rm vir}},
\end{equation}
where $v_{\rm rad}$ and $m_{\rm p}$ are the radial velocities and masses of individual gas particles (within 10 percent of the virial radius), respectively, and $R_{\rm vir}$ is the virial radius. Note that we assume that gas particles with $v_{\rm rad} < 0$ are being accreted onto the galaxy.

We see that galaxies that end up with weaker gradients grow steadily over cosmic time and experience significant cold gas accretion in the past 4-6 Gyrs. On the other hand, galaxies with stronger gradients set their sizes and accrete gas earlier than their weak-gradient counterparts.

\label{sec:origin}

%%%%%%%%%%%%%%%%%%%%%%%%%%%%%%%%%%%%%%%%%%%%%%%%%%%%%%%%%%%%%%%%%%%%%%%%%%%%%%%%%%%%%%%%%%%%%%%%%%%%%%
%%%%%%%%%%%%%%%%%%%%%%%%%%%%%%%% GAS-PHASE METALLICITY %%%%%%%%%%%%%%%%%%%%%%%%%%%%%%%%%%%%%%%%%%%%
%%%%%%%%%%%%%%%%%%%%%%%%%%%%%%%%%%%%%%%%%%%%%%%%%%%%%%%%%%%%%%%%%%%%%%%%%%%%%%%%%%%%%%%%%%%%%%%%%%%%%%

\subsection{Gas-Phase Metallicity Gradients}

\begin{figure}
	\includegraphics[width=\columnwidth, height=0.32
	\textheight,, trim = 0 0 0 0]{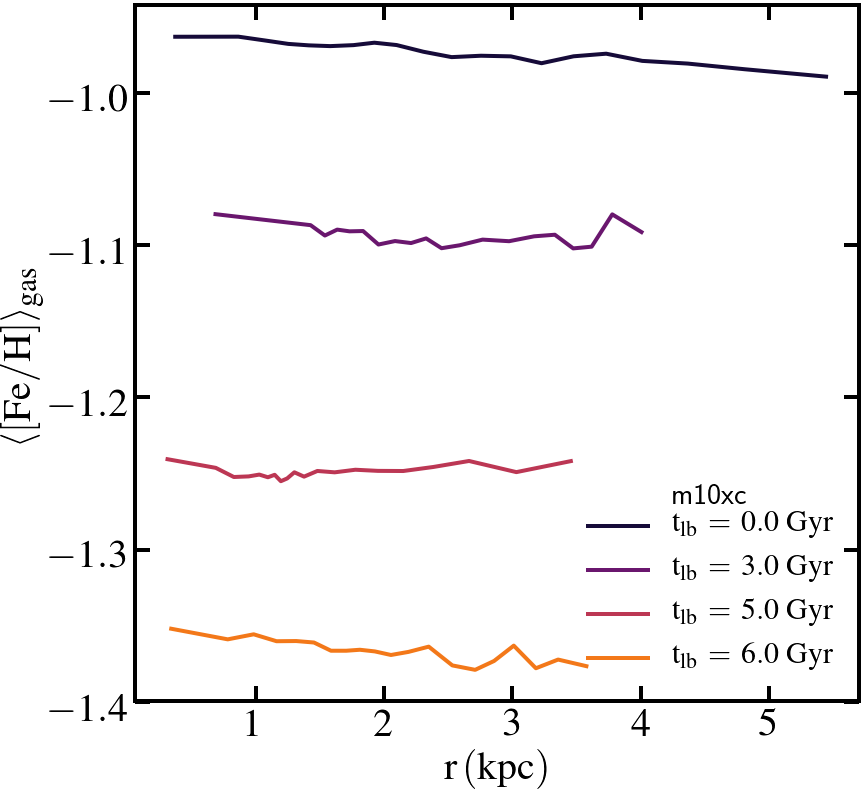}
	\centering
	\caption[migration all]{--- \textit{\textbf{Evolution of gas-phase metallicity profiles}.} The average gas-phase metallicity as a function of radius out to the 90 percent gas mass radius, within 0.1$R_{\rm vir}$, (for m10xc) at different points in time. The lines are colour-coded by lookback time. The gaseous component grows and becomes more enriched over time whilst simultaneously maintaining a relatively flat radial metallicity profile.}
	\label{fig:timegrad}
\end{figure}

Recall that in the bottom panel of Figure \ref{fig:migration} we showed that young stars at fixed lookback times tend to be distributed with fairly uniform metallicities, suggesting the lack of pre-existing gas-phase metallicity gradients. In this subsection, we explore gas-phase gradients in our simulated galaxies directly, and use these to develop a more complete picture of how stellar gradients arise. We also demonstrate that our simulations exhibit no correlation between stellar gradients and gas gradients at $z=0$.

Figure \ref{fig:timegrad} shows the evolution of the gas-phase metallicity gradient for one example galaxy, m10xc, at various lookback times. The lines extend out to the radius that contains 90 percent of the gas mass existing within 0.1$R_{\rm vir}$. We constrain our gas budget to temperatures cooler than $10^4$ K. Each profile is colour-coded by the lookback time. There are three results here. First, the extent of the gas component grows with time. Second, the average metallicity of the gas increases as a function of time. Finally, at fixed time, the metalliticy falls off weakly with radius. This result is consistent with our previous findings. Stars forming from gas at 6 Gyr lookback time would be both concentrated and metal poor.  Such stars would be prone to feedback puffing. Stars forming at very late times would trace extended, metal rich gas. We see this behaviour in gas-phase metallicity evolution in other galaxies within our sample in ways that are consistent with the broader picture painted above.

\begin{figure}
	\includegraphics[width=\columnwidth, height=0.32
	\textheight,, trim = 0 0 0 0]{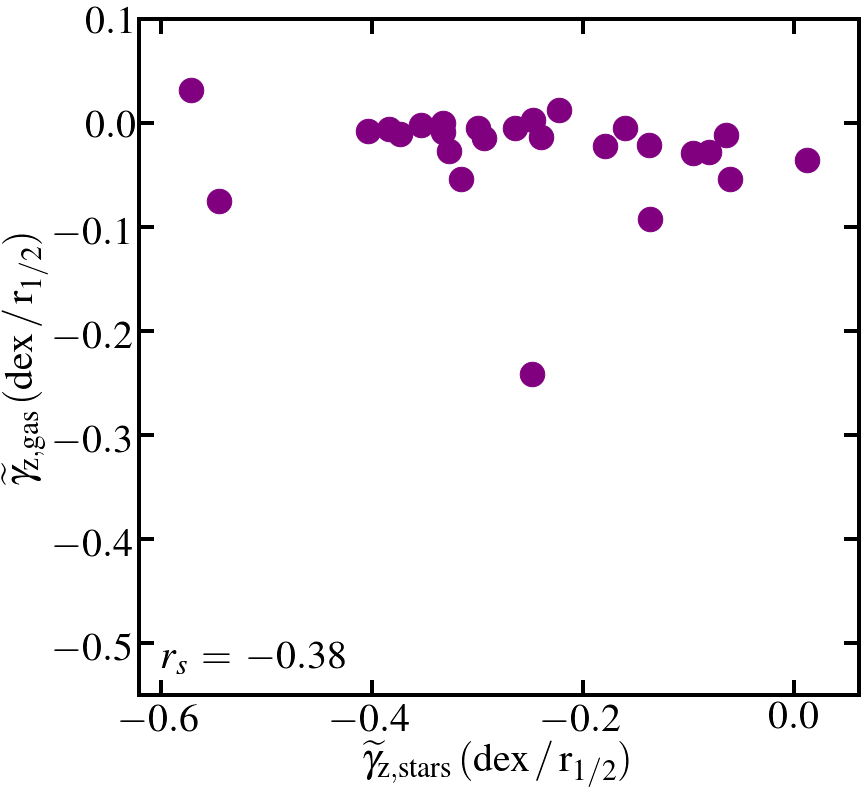}
	\centering
	\caption[migration all]{-- \textit{\textbf{Gas-phase versus stellar metallicity gradients}.} The strength of 3D gas-phase metallicity gradients versus 3D stellar metallicity gradients for the galaxies in our sample. Most of the galaxies have relatively flat gas-phase metallicity gradients but span a wide range of stellar metallicity gradients. We observe no clear relationship between the two sets ($r_{\rm s} \, = \, 0.38 $). Note that the steepest stellar metallicty gradients have quite flat gas-phase metallicity profiles.}
	\label{fig:gradvgrad}
\end{figure}

\begin{figure*}
	\includegraphics[width=1.0\textwidth,height=0.64
	\textheight, trim = 0 0 0 0]{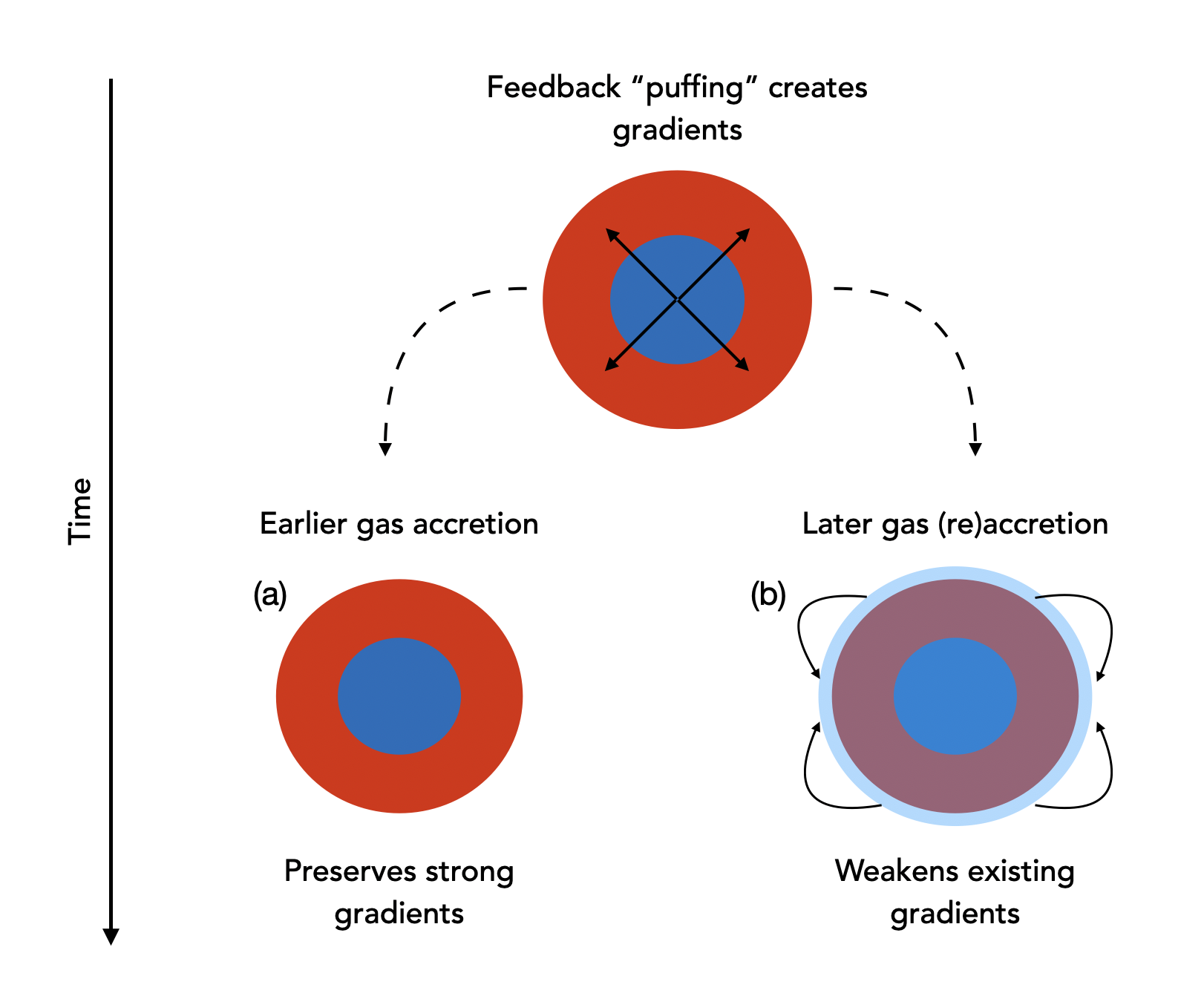}
	\hspace{.1in}
	\centering
	\caption[observations]{--- \textit{\textbf{The formation and evolution of stellar metallicity gradients.}} Negative stellar metallicity gradients in most FIRE-2 dwarf galaxies are formed as a result of periodic feedback events that preferentially puff older, more metal-poor stellar populations (red-shaded circles) outward. Younger, more metal-rich stellar populations (blue-shaded circles) live through fewer of these cycles and are puffed less. Subsequently, a galaxy's gradient strength can be set during one of two scenarios: (a) some galaxies tend to set their sizes and experience more gas accretion earlier on than do their weak-gradient counterparts. This serves to preserve strong, negative, stellar metallicity gradients. On the other hand, (b) some galaxies tend to grow steadily over time and experience significant gas accretion at late times. This gas is pre-enriched and tends to settle at large radius. This drives extended, late time star formation that is relatively metal rich and this that works to weaken/flatten existing stellar metallicity gradients.}
	\label{fig:diagram}
\end{figure*}

\begin{figure*}
	\includegraphics[width=1.0\textwidth,height=0.31
	\textheight, trim = 0 0 0 0]{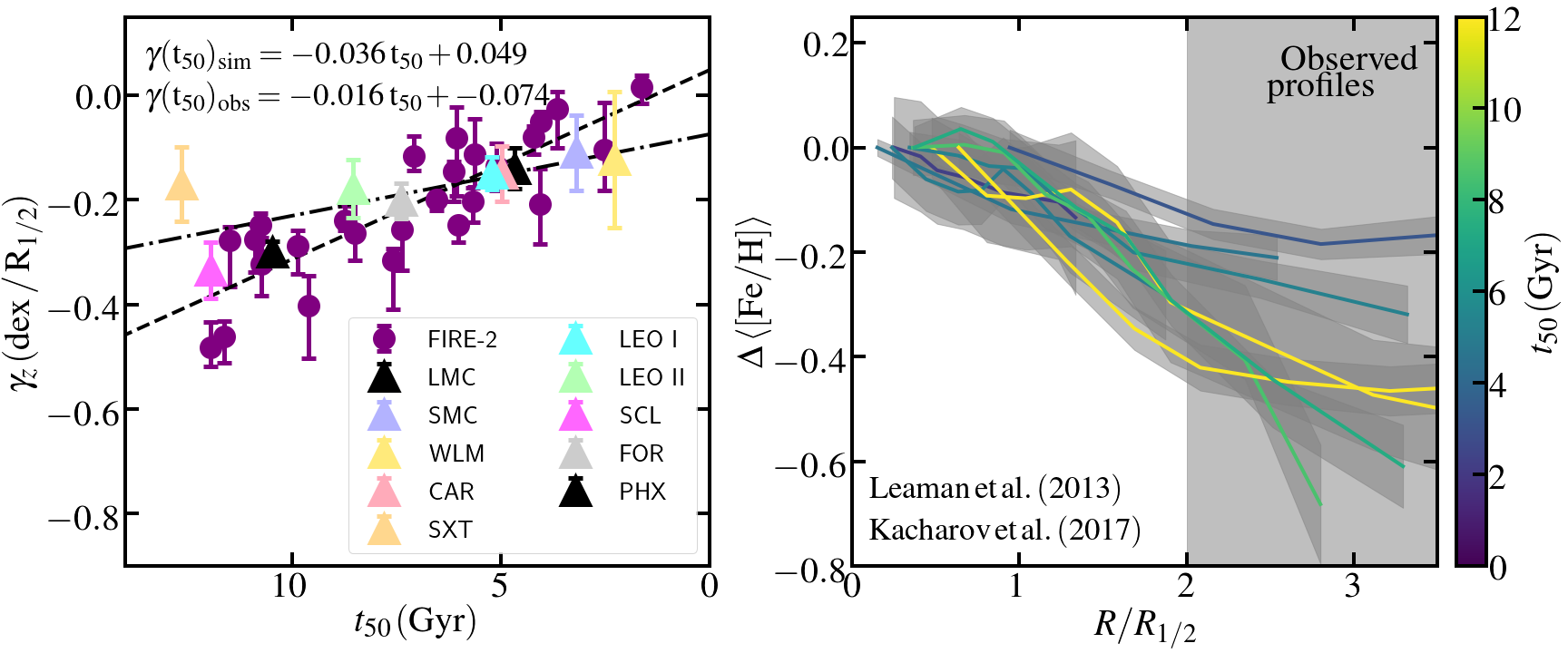}
	\hspace{.1in}
	\centering
	\caption[observations]{--- \textit{\textbf{An observed gradient-strength-galaxy-age relationship.}} A comparison between the simulated and observed gradient-strength-galaxy-age relationships. {\em Left:} Gradient strength versus median stellar age ($t_{50}$) of both the simulated (purple circles) and observed (pastel triangles) galaxy samples. The black, dashed and dash-dotted lines represent the least squares fit to the simulated and observed relationships, respectively. The corresponding fits are shown at the top left. While the simulated and observed galaxies follow similar gradient-strength-galaxy-age relationships, the observed sample seems to follow a slightly shallower relationship. {\em Right:} The average iron abundance for each observed galaxy as a function of projected radius in units of the $R_{1/2}$. The right panel can be compared to the theoretical models in Figure \ref{fig:all_grad}, where we also normalise the profiles to the metallicity at the centre of each galaxy and colour-coded by $t_{50}$ values. The shaded bands represent the $1\sigma$ uncertainty in the running averages. Similarly to the simulated sample we use the stellar metallicty data within $2R_{1/2}$ to determine the metallicity gradeint strength and thus exclude the data that extends into the shaded region.}
	\label{fig:obs_compare}
\end{figure*}

The relatively flat gas-phase metallicty gradient we see at $z=0$ in Figure \ref{fig:timegrad} is typical of our entire sample. This is illustrated in Figure \ref{fig:gradvgrad}, which shows the relationship between our galaxies' gas-phase and stellar metallicity gradient strengths at $z=0$. We determine the gas-phase metallicity gradients by employing the same approach as for their stellar counterparts. Whilst the galaxies exhibit a wide range of stellar metallicity gradients, their corresponding gas-phase gradients are all relatively flat and show no clear correlation with stellar metallicity gradients. There is only one galaxy, m10xh, where the gas-phase metallicty gradient is comparable to the stellar metallicity gradients. Whilst this galaxy is unique in that it has the most massive gaseous component of all the galaxies in our sample, more work must be done to determine the origins of its steep gas-phase metallicity gradient. However, whilst the inner gas-phase gradient is steep, the gradient is relatively flat when measured out to larger radii.

The results from Figures 8 \& 9 are in agreement with previous results based on the FIRE simulations, which show that gas in simulated galaxies at these mass scales is well mixed at all times \citep{Ma2017, Escala18}. The growth in the gas component, coupled with the metal-enrichment over time, further suggests that gas accretion and self enrichment (via the baryon cycle) play a key role in shaping their present-day stellar metallicity gradients.

\label{sec:gas}

\subsection{Summary explanation: why do stellar metallicty gradients correlate with galaxy age?}

Figures 4-8, together with previous work by \citet{ElBadry16}, motivate the following explanation for the trend we see between metallicity gradient strength and galaxy age (illustrated in Figure \ref{fig:diagram}): negative metallicity gradients in the FIRE-2 simulations are driven in dwarf galaxies by periodic feedback events that steadily puff older stars outward over time. Younger stars experience fewer feedback cycles and therefore experience less radial migration. Some galaxies experience significant gas accretion at late times, and this acts to flatten gradients. The late-time accretion is metal-enriched, tends to be deposited at larger radii because it has higher angular momentum \citep{Stewart2013,ElBadry2018, Grand2019}, and provides fuel for younger, radially-extended, metal-rich star formation. Previous analyses employing the FIRE simulations find that the re-accretion of gas previously ejected in galactic winds dominates late-time accretion, especially at the dwarf mass scale \citep{Angles2017}. Other galaxies experience earlier accretion of lower angular momentum gas, so they preserve steeper gradients driven by feedback puffing. The tendency for late-time gas accretion to flatten gradients is also seen in Milky-Way scale simulations by \citet{Grand2019}. Our findings are also in line with results from earlier literature, which suggest that secular processes -- such as gas accretion and feedback -- play a key role in shaping stellar metallicity gradients \citep{benitez2016, Revaz2018}.

%%%%%%%%%%%%%%%%%%%%%%%%%%%%%%%%%%%%%%%%%%%%%%%%%%%%%%%%%%%%%%%%%%%%%%%%%%%%%%%%%%%%%%%%%%%%%%%%%%%%%%
%%%%%%%%%%%%%%%%%%%%%%%%%%%%%%%% OBSERVATIONAL COMPARISON %%%%%%%%%%%%%%%%%%%%%%%%%%%%%%%%%%%%%%%%%%%%
%%%%%%%%%%%%%%%%%%%%%%%%%%%%%%%%%%%%%%%%%%%%%%%%%%%%%%%%%%%%%%%%%%%%%%%%%%%%%%%%%%%%%%%%%%%%%%%%%%%%%%
\section{Observational Comparison}

For our comparison with observations we use the results of \citet{Leaman13} and \citet{Kacharov17} to compile stellar metallicity gradient information for a total of 10 Local Group dwarfs. We then use $t_{50}$ values determined from the star formation histories created in \citet{Weisz2014a, Weisz2014b}, and \citet{Bettinelli18}. Table \ref{table:obs_galaxies} summarises the $t_{50}$ values, gradient strengths, $\gamma_{z}$, and the $R_{1/2}$ values taken from \citet{McConnachie12}, \citet{Munoz18}, and \citet{Simon19} -- for the ten galaxies we analyse. Note that the observed galaxies' gradient strengths and associated errors are determined by bootstrapped sampling the observed profile data. The quoted error in the gradient strengths represent the  $1 \sigma$ spread about the average. We define $\gamma_z$ the same way we have defined it in our simulated measurements, using the slope of a least-squares fit from $R = 0$ to $R = 2 R_{1/2}$.

\begin{table}
  \caption{Properties of 10 observed Local Group galaxies: (1) lookback time to the formation of 50\% of stars determined from published SFHs \citep{Weisz2014a, Weisz2014b, Bettinelli18}, (2) calculated stellar metallicity gradients including error bars, (3) galaxy half-light radius}
    \centering % used for centering table
        %\begin{tabularx}{\textwidth}{lccccc}
        \begin{tabular}{cccc}
            \hline
            \hline  %inserts double horizontal lines
            $\rm Galaxy$  & \tfifty & $\gamma_{z}$ & $R_{1/2}$\\
            $\rm Name$ & [Gyr] & $[dex/R_{1/2}]$ & [kpc]\\
      
            \hline
            MW Dwarfs & (1) & (2) & (3)\\
            \hline

            WLM        &    2.28    &   -0.11 $\pm$  0.12   &  2.111\\
            SMC        &    3.21    &   -0.13 $\pm$  0.07   &  1.106\\
            LMC        &    4.68    &   -0.15 $\pm$  0.04   &  2.697\\
            Carina     &    4.99    &   -0.16 $\pm$  0.05   &  0.311\\
            Leo I      &    5.22    &   -0.15 $\pm$  0.02   &  0.270\\
            Fornax     &    7.40    &   -0.21 $\pm$  0.03   &  0.792\\
            Leo II     &    8.54    &   -0.18 $\pm$  0.06   &  0.171\\
            Phoenix    &    10.48   &   -0.30 $\pm$  0.02   &  0.454\\
            Sculptor   &    11.95   &   -0.33 $\pm$  0.06   &  0.279\\
            Sextans    &    12.64   &   -0.16 $\pm$  0.07   &  0.456\\
            
            \hline
            \hline
            \label{table:obs_galaxies}
        \end{tabular}
\end{table}

The left panel of Figure \ref{fig:obs_compare} compares the simulated and observed gradient-strength-galaxy-age relationship. Purple circles show simulated galaxies, whilst pastel-coloured triangles represent observed Local Group galaxies (galaxy names indicated). On the right, we present the mean metallicity profiles of the observed galaxies as a function of projected radius in units of $R_{1/2}$. As in Figure \ref{fig:all_grad}, which shows the same for our simulated galaxies, we normalise the metallicity profiles to the values at the centre of each galaxy. The solid lines depict the average metallicity at a given radius. The colour corresponds to each galaxy's median age, as indicated by the colour bar. The grey bands represent the $1 \sigma$ uncertainty of the running average stellar metallicity \citep[calculated by][]{Leaman13}. 

Similarly to the simulated galaxy sample, the observed galaxy sample follows a gradient-strength-galaxy-age relationship such that galaxies with older (younger) stellar populations tend to have stronger (weaker) stellar metallicity gradients. This is intriguing because most of the observed galaxies on this figure are satellites of the Milky Way, whilst our simulations are of isolated systems. This suggests that internal star formation processes play a dominant role in creating stellar metallicity gradients with a diversity of strengths -- thus implying that dynamical/environmental effects, such as ram pressure stripping, play a more limited, secondary role in shaping metallicity gradients by limiting late-time gas accretion, thereby preventing the flattening effect we see in our youngest galaxies (see previous section). The observed sample of galaxies follow this relationship: 
\begin{equation}
    \gamma(t_{50})_{\rm obs} = -0.015 t_{50} + 0.081,
\end{equation}
The slope for this observed relationship is $-0.015 \pm 0.005$ whilst the slope of the relationship followed by the simulated sample is $-0.036 \pm 0.005$. The apparent disagreement in the slopes suggests that whilst the observed galaxies follow a similar gradient-strength-galaxy-age relationship to the simulated sample, the slope is slightly flatter (relationship represented by the dash-dotted line). It is possible that the strong feedback in the FIRE-2 implementation can lead to stellar metallicity gradients that are stronger than what we would expect to see in the real universe - resulting in a steeper gradient-strength-galaxy-age relationship. Determining the ages and stellar metallicity gradients for more dwarf galaxies will help us determine the extent of this discrepancy between the simulated and observed galaxies.

One clear outlier in the left panel of Figure \ref{fig:obs_compare} is Sextans (pink triangle). This galaxy has a fairly flat gradient compared to simulated galaxies of similar (old) age. The metallicity profile of Sextans is somewhat unusual, as can be seen by the yellow line in the right panel with a fairly flat slope out to $1.5R_{1/2}$ accompanied by a steep gradient beyond that. In keeping with the $\gamma_z$ definition used elsewhere, we have defined it using the slope out to $2R_{1/2}$, but had we defined it out to larger radii we would have measured a steeper slope, more in line with the global trend.

\label{sec:obs}
%%%%%%%%%%%%%%%%%%%%%%%%%%%%%%%%%%%%%%%%%%%%%%%%%%%%%%%%%%%%%%%%%%%%%%%%%%%%%%%%%
%%%%%%%%%%%%%%%%%%%%%%%%%%%%%%%% CONCLUSION %%%%%%%%%%%%%%%%%%%%%%%%%%%%%%%%%%%%%
%%%%%%%%%%%%%%%%%%%%%%%%%%%%%%%%%%%%%%%%%%%%%%%%%%%%%%%%%%%%%%%%%%%%%%%%%%%%%%%%%

\section{Conclusions}

In this paper we examine 26 simulated dwarf galaxies using FIRE-2 physics \citep{Hopkins18} to explore the origin and nature of radial, stellar metallicity gradients. The galaxies have stellar masses between $10^{5.5}$ and $10^{8.6} \msun$ and $M_{\rm vir} \simeq$ $10^{10}\msun$ -- $10^{11}\msun$. Most of the galaxies in this sample exhibit negative stellar metallicity gradients, with interiors more metal-rich than the outskirts. We predict a correlation between the stellar metallicity gradient strength and overall galaxy age, as measured by the median stellar age of the galaxy at $z=0$: $\gamma(t_{50}) = -0.036 t_{50} + 0.049$. Older galaxies tend to have stronger gradients. \citet{Graus19} finds a correlation between galaxy age and age gradient using the same galaxy sample. whilst these galaxies exhibit negative stellar metallicity gradients, it is important to note that these galaxies have well mixed gas-phase metallicities (and thus, do not exhibit a gas-phase gradient) at any given point in time \citep{Ma2017, Escala18}.

By studying the evolution of stars and gas with time in these systems, we conclude that strong negative stellar metallicity gradients arise from the steady ``puffing'' effects of feedback, which tend to drive the oldest, most metal-poor stars outward over time (Figure \ref{fig:migration_all}). This mechanism, detailed by \citet{ElBadry16}, is a result of stars gravitationally reacting to gas outflows occurring in conjunction with repeated starburst events. Although feedback puffing appears to be a universal feature amongst our galaxies, there is a fair amount of variance in the level of late-time gas accretion. Recent accretion tends to flatten out gradients. This is because gas that is deposited at late times tends to be recycled, enriched, and deposited at large radii \citep[see][and also our Figure \ref{fig:timegrad}]{Angles2017}. Stars that form out of this late-accreted gas create an extended, metal-rich stellar population, which washes out any previous gradient set by the radial migration outward of old, metal-poor stars. This overall picture is sketched in Figure \ref{fig:diagram}.

Using published data from 10 Local Group dwarfs \citep{Leaman13, Kacharov17}, we show that they do appear to follow the gradient-strength-galaxy-age relationship predicted by our simulations (Figure \ref{fig:obs_compare}). This suggests that stellar metallicity gradients in real galaxies may be largely governed by a competition between feedback-puffing of old/metal-poor stars and late-time star formation from recently accreted recycled/metal-enriched gas. We conclude that our prediction that dwarf galaxy stellar metallicity gradient strength should correlate with galaxy age is consistent with current observations, though the slope of the correlation appears to be flatter in the observed population. This suggests that internal feedback mechanisms and associated baryon-cycle enrichment may play the dominant role in driving stellar metallicity gradients in dwarf galaxies rather than environmental factors.

Future work exploring the existence and strength of any age/gradient relationships in other simulation codes may provide an avenue for testing feedback models. Similarly, larger, more complete observational samples from future telescopes like WFIRST and JWST may enable more detailed and quantitative comparisons.

\label{sec:Conclusion}

%%%%%%%%%%%%%%%%%%%%%%%%%%%%%%%%%%%%%%%%%%%%%%%%%%%%%%%%%%%%%%%%%%%%%%%%%%%%%%%%%
%%%%%%%%%%%%%%%%%%%%%%%%%%%%%%%% ACKNOWLEDGMENTS %%%%%%%%%%%%%%%%%%%%%%%%%%%%%%%%%%%%%
%%%%%%%%%%%%%%%%%%%%%%%%%%%%%%%%%%%%%%%%%%%%%%%%%%%%%%%%%%%%%%%%%%%%%%%%%%%%%%%%%

\section{Acknowledgments}
We dedicate this paper to Perla Maritza Mercado and José Antonio Florez Velázquez -- two bright lights that had a tremendous impact on FJM's life. \textit{Que en paz descansen}.

We honour the invaluable labour of the maintenance and clerical staff at our institutions, whose contributions make our scientific discoveries a reality. This research was conducted on Acjachemen and Tongva Indigenous land.

The functionalities provided by the NumPy library played a critical role in the analysis presented in this paper \citep{vanderWalt2011}. We used the FIRE studio (\href{https://github.com/agurvich/FIRE_studio}{learn more here}), an open source Python visualization package designed with the FIRE simulations in mind, to create the images in Figure \ref{fig:mock_hubble}. We used the WebPlotDigitizer tool (\href{https://apps.automeris.io/wpd/}{https://apps.automeris.io/wpd/}) to determine the median stellar age, $t_{50}$, from the 10 observed galaxies' star formation histories. 

FJM and JSB were supported by NSF Grants AST-1518291 and AST-1910965. We thank Glenn Tipton and Bill Ward for inspiration. MBK acknowledges support from NSF CAREER award AST-1752913, NSF grant AST-1910346, NASA grant NNX17AG29G, and HST-AR-15006, HST-AR-15809, HST-GO-15658, HST-GO-15901, and HST-GO-15902 from the Space Telescope Science Institute, which is operated by AURA, Inc., under NASA contract NAS5-26555. AW received support from NASA through ATP grant 80NSSC18K1097 and HST grants GO-14734, AR-15057, AR-15809, and GO-15902 from STScI; the Heising-Simons Foundation; and a Hellman Fellowship. Support for JM is provided by the
NSF (AST Award Number 1516374), and by the Harvard Institute for Theory and Computation, through their Visiting Scholars Program. ASG is supported by the Harlan J. Smith postdoctoral fellowship. CAFG was supported by NSF through grants AST-1715216 and CAREER award AST-1652522; by NASA through grant 17-ATP17-0067; and by a Cottrell Scholar Award from the Research Corporation for Science Advancement.

\section{Data Availability}
The data supporting the plots within this article are available on reasonable request to the corresponding author. A public version of the GIZMO code is available at \href{http://www.tapir.caltech.edu/~phopkins/Site/GIZMO.html}{http://www.tapir.caltech.edu/~phopkins/Site/GIZMO.html}. Additional data including simulation snapshots, initial conditions, and derived data products are available at \href{https://fire.northwestern.edu/data/}{https://fire.northwestern.edu/data/}.

%%%%%%%%%%%%%%%%%%%%%%%%%%%%%%%%%%%%%%%%%%%%%%%%%%%%%%%%%%%%%%%%%%%%%%%%%%%%%%%%%
%%%%%%%%%%%%%%%%%%%%%%%%%%%%%% FIGURES FOR NOW %%%%%%%%%%%%%%%%%%%%%%%%%%%%%%%%%%
%%%%%%%%%%%%%%%%%%%%%%%%%%%%%%%%%%%%%%%%%%%%%%%%%%%%%%%%%%%%%%%%%%%%%%%%%%%%%%%%%
% Will have to move to correct positions later 

%%%%%%%%%%%%%%%%%%%%%%%%%%%%%%%%%%%%%%%%%%%%%%%%%%



\bibliographystyle{mnras}
\bibliography{Grad_paper_refs.bib}

\begin{thebibliography}{}
\makeatletter
\relax
\def\mn@urlcharsother{\let\do\@makeother \do\$\do\&\do\#\do\^\do\_\do\%\do\~}
\def\mn@doi{\begingroup\mn@urlcharsother \@ifnextchar [ {\mn@doi@}
  {\mn@doi@[]}}
\def\mn@doi@[#1]#2{\def\@tempa{#1}\ifx\@tempa\@empty \href
  {http://dx.doi.org/#2} {doi:#2}\else \href {http://dx.doi.org/#2} {#1}\fi
  \endgroup}
\def\mn@eprint#1#2{\mn@eprint@#1:#2::\@nil}
\def\mn@eprint@arXiv#1{\href {http://arxiv.org/abs/#1} {{\tt arXiv:#1}}}
\def\mn@eprint@dblp#1{\href {http://dblp.uni-trier.de/rec/bibtex/#1.xml}
  {dblp:#1}}
\def\mn@eprint@#1:#2:#3:#4\@nil{\def\@tempa {#1}\def\@tempb {#2}\def\@tempc
  {#3}\ifx \@tempc \@empty \let \@tempc \@tempb \let \@tempb \@tempa \fi \ifx
  \@tempb \@empty \def\@tempb {arXiv}\fi \@ifundefined
  {mn@eprint@\@tempb}{\@tempb:\@tempc}{\expandafter \expandafter \csname
  mn@eprint@\@tempb\endcsname \expandafter{\@tempc}}}

\bibitem[\protect\citeauthoryear{{Angl{\'e}s-Alc{\'a}zar},
  {Faucher-Gigu{\`e}re}, {Kere{\v{s}}}, {Hopkins}, {Quataert}  \&
  {Murray}}{{Angl{\'e}s-Alc{\'a}zar} et~al.}{2017}]{Angles2017}
{Angl{\'e}s-Alc{\'a}zar} D.,  {Faucher-Gigu{\`e}re} C.-A.,  {Kere{\v{s}}} D.,
  {Hopkins} P.~F.,  {Quataert} E.,   {Murray} N.,  2017, \mn@doi [\mnras]
  {10.1093/mnras/stx1517}, \href
  {https://ui.adsabs.harvard.edu/abs/2017MNRAS.470.4698A} {470, 4698}

\bibitem[\protect\citeauthoryear{{Battaglia} et~al.,}{{Battaglia}
  et~al.}{2006}]{Battaglia06}
{Battaglia} G.,  et~al., 2006, \mn@doi [\aap] {10.1051/0004-6361:20065720},
  \href {https://ui.adsabs.harvard.edu/abs/2006A&A...459..423B} {459, 423}

\bibitem[\protect\citeauthoryear{{Battaglia}, {Tolstoy}, {Helmi}, {Irwin},
  {Parisi}, {Hill}  \& {Jablonka}}{{Battaglia} et~al.}{2011}]{Battaglia11}
{Battaglia} G.,  {Tolstoy} E.,  {Helmi} A.,  {Irwin} M.,  {Parisi} P.,  {Hill}
  V.,   {Jablonka} P.,  2011, \mn@doi [\mnras]
  {10.1111/j.1365-2966.2010.17745.x}, \href
  {https://ui.adsabs.harvard.edu/abs/2011MNRAS.411.1013B} {411, 1013}

\bibitem[\protect\citeauthoryear{{Ben{\'\i}tez-Llambay}, {Navarro}, {Abadi},
  {Gottl{\"o}ber}, {Yepes}, {Hoffman}  \& {Steinmetz}}{{Ben{\'\i}tez-Llambay}
  et~al.}{2016}]{benitez2016}
{Ben{\'\i}tez-Llambay} A.,  {Navarro} J.~F.,  {Abadi} M.~G.,  {Gottl{\"o}ber}
  S.,  {Yepes} G.,  {Hoffman} Y.,   {Steinmetz} M.,  2016, \mn@doi [\mnras]
  {10.1093/mnras/stv2722}, \href
  {https://ui.adsabs.harvard.edu/abs/2016MNRAS.456.1185B} {456, 1185}

\bibitem[\protect\citeauthoryear{{Bettinelli}, {Hidalgo}, {Cassisi}, {Aparicio}
   \& {Piotto}}{{Bettinelli} et~al.}{2018}]{Bettinelli18}
{Bettinelli} M.,  {Hidalgo} S.~L.,  {Cassisi} S.,  {Aparicio} A.,   {Piotto}
  G.,  2018, \mn@doi [\mnras] {10.1093/mnras/sty226}, \href
  {https://ui.adsabs.harvard.edu/abs/2018MNRAS.476...71B} {476, 71}

\bibitem[\protect\citeauthoryear{{Chan}, {Kere{\v{s}}}, {O{\~n}orbe},
  {Hopkins}, {Muratov}, {Faucher-Gigu{\`e}re}  \& {Quataert}}{{Chan}
  et~al.}{2015}]{Chan2015}
{Chan} T.~K.,  {Kere{\v{s}}} D.,  {O{\~n}orbe} J.,  {Hopkins} P.~F.,  {Muratov}
  A.~L.,  {Faucher-Gigu{\`e}re} C.~A.,   {Quataert} E.,  2015, \mn@doi [\mnras]
  {10.1093/mnras/stv2165}, \href
  {https://ui.adsabs.harvard.edu/abs/2015MNRAS.454.2981C} {454, 2981}

\bibitem[\protect\citeauthoryear{{De Young} \& {Heckman}}{{De Young} \&
  {Heckman}}{1994}]{De94}
{De Young} D.~S.,  {Heckman} T.~M.,  1994, \mn@doi [\apj] {10.1086/174510},
  \href {https://ui.adsabs.harvard.edu/abs/1994ApJ...431..598D} {431, 598}

\bibitem[\protect\citeauthoryear{{Dekel} \& {Silk}}{{Dekel} \&
  {Silk}}{1986}]{DekelSilk86}
{Dekel} A.,  {Silk} J.,  1986, \mn@doi [\apj] {10.1086/164050}, \href
  {https://ui.adsabs.harvard.edu/abs/1986ApJ...303...39D} {303, 39}

\bibitem[\protect\citeauthoryear{{El-Badry}, {Wetzel}, {Geha}, {Hopkins},
  {Kere{\v s}}, {Chan}  \& {Faucher-Gigu{\`e}re}}{{El-Badry}
  et~al.}{2016}]{ElBadry16}
{El-Badry} K.,  {Wetzel} A.,  {Geha} M.,  {Hopkins} P.~F.,  {Kere{\v s}} D.,
  {Chan} T.~K.,   {Faucher-Gigu{\`e}re} C.-A.,  2016, \mn@doi [\apj]
  {10.3847/0004-637X/820/2/131}, \href
  {http://adsabs.harvard.edu/abs/2016ApJ...820..131E} {820, 131}

\bibitem[\protect\citeauthoryear{{El-Badry} et~al.,}{{El-Badry}
  et~al.}{2018}]{ElBadry2018}
{El-Badry} K.,  et~al., 2018, \mn@doi [\mnras] {10.1093/mnras/stx2482}, \href
  {https://ui.adsabs.harvard.edu/abs/2018MNRAS.473.1930E} {473, 1930}

\bibitem[\protect\citeauthoryear{{Escala} et~al.,}{{Escala}
  et~al.}{2018}]{Escala18}
{Escala} I.,  et~al., 2018, \mn@doi [\mnras] {10.1093/mnras/stx2858}, \href
  {https://ui.adsabs.harvard.edu/abs/2018MNRAS.474.2194E} {474, 2194}

\bibitem[\protect\citeauthoryear{{Faucher-Gigu{\`e}re}}{{Faucher-Gigu{\`e}re}}{2018}]{FG2018}
{Faucher-Gigu{\`e}re} C.-A.,  2018, \mn@doi [\mnras] {10.1093/mnras/stx2595},
  \href {https://ui.adsabs.harvard.edu/abs/2018MNRAS.473.3717F} {473, 3717}

\bibitem[\protect\citeauthoryear{{Faucher-Gigu{\`e}re}, {Lidz}, {Zaldarriaga}
  \& {Hernquist}}{{Faucher-Gigu{\`e}re} et~al.}{2009}]{FG2009}
{Faucher-Gigu{\`e}re} C.-A.,  {Lidz} A.,  {Zaldarriaga} M.,   {Hernquist} L.,
  2009, \mn@doi [\apj] {10.1088/0004-637X/703/2/1416}, \href
  {https://ui.adsabs.harvard.edu/abs/2009ApJ...703.1416F} {703, 1416}

\bibitem[\protect\citeauthoryear{{Faucher-Gigu{\`e}re}, {Kere{\v{s}}}  \&
  {Ma}}{{Faucher-Gigu{\`e}re} et~al.}{2011}]{FG2011}
{Faucher-Gigu{\`e}re} C.-A.,  {Kere{\v{s}}} D.,   {Ma} C.-P.,  2011, \mn@doi
  [\mnras] {10.1111/j.1365-2966.2011.19457.x}, \href
  {https://ui.adsabs.harvard.edu/abs/2011MNRAS.417.2982F} {417, 2982}

\bibitem[\protect\citeauthoryear{{Fitts} et~al.,}{{Fitts}
  et~al.}{2017}]{Fitts17}
{Fitts} A.,  et~al., 2017, \mn@doi [\mnras] {10.1093/mnras/stx1757}, \href
  {http://adsabs.harvard.edu/abs/2017MNRAS.471.3547F} {471, 3547}

\bibitem[\protect\citeauthoryear{{Fitts} et~al.,}{{Fitts}
  et~al.}{2018}]{Fitts18}
{Fitts} A.,  et~al., 2018, preprint, \href
  {http://adsabs.harvard.edu/abs/2018arXiv180106187F} {} (\mn@eprint {arXiv}
  {1801.06187})

\bibitem[\protect\citeauthoryear{{Flores Vel{\'a}zquez} et~al.,}{{Flores
  Vel{\'a}zquez} et~al.}{2020}]{FV2020}
{Flores Vel{\'a}zquez} J.~A.,  et~al., 2020, arXiv e-prints, \href
  {https://ui.adsabs.harvard.edu/abs/2020arXiv200808582F} {p. arXiv:2008.08582}

\bibitem[\protect\citeauthoryear{{Grand} et~al.,}{{Grand}
  et~al.}{2019}]{Grand2019}
{Grand} R. J.~J.,  et~al., 2019, \mn@doi [\mnras] {10.1093/mnras/stz2928},
  \href {https://ui.adsabs.harvard.edu/abs/2019MNRAS.490.4786G} {490, 4786}

\bibitem[\protect\citeauthoryear{{Graus} et~al.,}{{Graus}
  et~al.}{2019}]{Graus19}
{Graus} A.~S.,  et~al., 2019, \mn@doi [\mnras] {10.1093/mnras/stz2649}, \href
  {https://ui.adsabs.harvard.edu/abs/2019MNRAS.490.1186G} {490, 1186}

\bibitem[\protect\citeauthoryear{{Grebel}}{{Grebel}}{1999}]{Grebel1999}
{Grebel} E.~K.,  1999, in {Whitelock} P.,  {Cannon} R.,  eds,  IAU Symposium
  Vol. 192, The Stellar Content of Local Group Galaxies. p.~17 (\mn@eprint
  {arXiv} {astro-ph/9812443})

\bibitem[\protect\citeauthoryear{{Harbeck} et~al.,}{{Harbeck}
  et~al.}{2001}]{Harbeck01}
{Harbeck} D.,  et~al., 2001, \mn@doi [\aj] {10.1086/324232}, \href
  {https://ui.adsabs.harvard.edu/abs/2001AJ....122.3092H} {122, 3092}

\bibitem[\protect\citeauthoryear{{Ho}, {Geha}, {Tollerud}, {Zinn},
  {Guhathakurta}  \& {Vargas}}{{Ho} et~al.}{2015}]{Ho15}
{Ho} N.,  {Geha} M.,  {Tollerud} E.~J.,  {Zinn} R.,  {Guhathakurta} P.,
  {Vargas} L.~C.,  2015, \mn@doi [\apj] {10.1088/0004-637X/798/2/77}, \href
  {https://ui.adsabs.harvard.edu/abs/2015ApJ...798...77H} {798, 77}

\bibitem[\protect\citeauthoryear{{Hopkins}}{{Hopkins}}{2015}]{Hopkins15}
{Hopkins} P.~F.,  2015, \mn@doi [\mnras] {10.1093/mnras/stv195}, \href
  {http://adsabs.harvard.edu/abs/2015MNRAS.450...53H} {450, 53}

\bibitem[\protect\citeauthoryear{{Hopkins} et~al.,}{{Hopkins}
  et~al.}{2018}]{Hopkins18}
{Hopkins} P.~F.,  et~al., 2018, \mn@doi [\mnras] {10.1093/mnras/sty1690}, \href
  {http://adsabs.harvard.edu/abs/2018MNRAS.480..800H} {480, 800}

\bibitem[\protect\citeauthoryear{{Kacharov} et~al.,}{{Kacharov}
  et~al.}{2017}]{Kacharov17}
{Kacharov} N.,  et~al., 2017, \mn@doi [\mnras] {10.1093/mnras/stw3188}, \href
  {https://ui.adsabs.harvard.edu/abs/2017MNRAS.466.2006K} {466, 2006}

\bibitem[\protect\citeauthoryear{{Kirby} et~al.,}{{Kirby}
  et~al.}{2010}]{Kirby10}
{Kirby} E.~N.,  et~al., 2010, \mn@doi [\apjs] {10.1088/0067-0049/191/2/352},
  \href {https://ui.adsabs.harvard.edu/abs/2010ApJS..191..352K} {191, 352}

\bibitem[\protect\citeauthoryear{{Kirby}, {Cohen}, {Guhathakurta}, {Cheng},
  {Bullock}  \& {Gallazzi}}{{Kirby} et~al.}{2013}]{Kirby13}
{Kirby} E.~N.,  {Cohen} J.~G.,  {Guhathakurta} P.,  {Cheng} L.,  {Bullock}
  J.~S.,   {Gallazzi} A.,  2013, \mn@doi [\apj] {10.1088/0004-637X/779/2/102},
  \href {https://ui.adsabs.harvard.edu/abs/2013ApJ...779..102K} {779, 102}

\bibitem[\protect\citeauthoryear{{Koch}, {Kleyna}, {Wilkinson}, {Grebel},
  {Gilmore}, {Evans}, {Wyse}  \& {Harbeck}}{{Koch} et~al.}{2007}]{Koch2007a}
{Koch} A.,  {Kleyna} J.~T.,  {Wilkinson} M.~I.,  {Grebel} E.~K.,  {Gilmore}
  G.~F.,  {Evans} N.~W.,  {Wyse} R. F.~G.,   {Harbeck} D.~R.,  2007, \mn@doi
  [\aj] {10.1086/519380}, \href
  {https://ui.adsabs.harvard.edu/abs/2007AJ....134..566K} {134, 566}

\bibitem[\protect\citeauthoryear{{Koleva}, {Prugniel}, {De Rijcke}  \&
  {Zeilinger}}{{Koleva} et~al.}{2011}]{Koleva11}
{Koleva} M.,  {Prugniel} P.,  {De Rijcke} S.,   {Zeilinger} W.~W.,  2011,
  \mn@doi [\mnras] {10.1111/j.1365-2966.2011.19057.x}, \href
  {https://ui.adsabs.harvard.edu/abs/2011MNRAS.417.1643K} {417, 1643}

\bibitem[\protect\citeauthoryear{{Kroupa}}{{Kroupa}}{2002}]{Kroupa02}
{Kroupa} P.,  2002, \mn@doi [Science] {10.1126/science.1067524}, \href
  {https://ui.adsabs.harvard.edu/abs/2002Sci...295...82K} {295, 82}

\bibitem[\protect\citeauthoryear{{Leaman} et~al.,}{{Leaman}
  et~al.}{2013}]{Leaman13}
{Leaman} R.,  et~al., 2013, \mn@doi [\apj] {10.1088/0004-637X/767/2/131}, \href
  {https://ui.adsabs.harvard.edu/abs/2013ApJ...767..131L} {767, 131}

\bibitem[\protect\citeauthoryear{{Leitherer} et~al.,}{{Leitherer}
  et~al.}{1999}]{Leitherer99}
{Leitherer} C.,  et~al., 1999, \mn@doi [\apjs] {10.1086/313233}, \href
  {https://ui.adsabs.harvard.edu/abs/1999ApJS..123....3L} {123, 3}

\bibitem[\protect\citeauthoryear{{Ma}, {Hopkins}, {Feldmann}, {Torrey},
  {Faucher-Gigu{\`e}re}  \& {Kere{\v{s}}}}{{Ma} et~al.}{2017}]{Ma2017}
{Ma} X.,  {Hopkins} P.~F.,  {Feldmann} R.,  {Torrey} P.,  {Faucher-Gigu{\`e}re}
  C.-A.,   {Kere{\v{s}}} D.,  2017, \mn@doi [\mnras] {10.1093/mnras/stx034},
  \href {https://ui.adsabs.harvard.edu/abs/2017MNRAS.466.4780M} {466, 4780}

\bibitem[\protect\citeauthoryear{{Martin} et~al.,}{{Martin}
  et~al.}{2020}]{Martin2020}
{Martin} G.,  et~al., 2020, arXiv e-prints, \href
  {https://ui.adsabs.harvard.edu/abs/2020arXiv200707913M} {p. arXiv:2007.07913}

\bibitem[\protect\citeauthoryear{{Mateo}}{{Mateo}}{1998}]{Mateo1998}
{Mateo} M.~L.,  1998, \mn@doi [\araa] {10.1146/annurev.astro.36.1.435}, \href
  {https://ui.adsabs.harvard.edu/abs/1998ARA&A..36..435M} {36, 435}

\bibitem[\protect\citeauthoryear{{Mateo}, {Olszewski}  \& {Walker}}{{Mateo}
  et~al.}{2008}]{Mateo2008}
{Mateo} M.,  {Olszewski} E.~W.,   {Walker} M.~G.,  2008, \mn@doi [\apj]
  {10.1086/522326}, \href
  {https://ui.adsabs.harvard.edu/abs/2008ApJ...675..201M} {675, 201}

\bibitem[\protect\citeauthoryear{{Mayer}, {Governato}, {Colpi}, {Moore},
  {Quinn}, {Wadsley}, {Stadel}  \& {Lake}}{{Mayer} et~al.}{2001}]{Mayer01}
{Mayer} L.,  {Governato} F.,  {Colpi} M.,  {Moore} B.,  {Quinn} T.,  {Wadsley}
  J.,  {Stadel} J.,   {Lake} G.,  2001, \mn@doi [\apj] {10.1086/322356}, \href
  {https://ui.adsabs.harvard.edu/abs/2001ApJ...559..754M} {559, 754}

\bibitem[\protect\citeauthoryear{{Mayer}, {Kazantzidis}, {Mastropietro}  \&
  {Wadsley}}{{Mayer} et~al.}{2007}]{Mayer07}
{Mayer} L.,  {Kazantzidis} S.,  {Mastropietro} C.,   {Wadsley} J.,  2007,
  \mn@doi [\nat] {10.1038/nature05552}, \href
  {https://ui.adsabs.harvard.edu/abs/2007Natur.445..738M} {445, 738}

\bibitem[\protect\citeauthoryear{{McConnachie}}{{McConnachie}}{2012}]{McConnachie12}
{McConnachie} A.~W.,  2012, \mn@doi [\aj] {10.1088/0004-6256/144/1/4}, \href
  {https://ui.adsabs.harvard.edu/abs/2012AJ....144....4M} {144, 4}

\bibitem[\protect\citeauthoryear{{Mu{\~n}oz} et~al.,}{{Mu{\~n}oz}
  et~al.}{2005}]{Munoz2005}
{Mu{\~n}oz} R.~R.,  et~al., 2005, \mn@doi [\apjl] {10.1086/497396}, \href
  {https://ui.adsabs.harvard.edu/abs/2005ApJ...631L.137M} {631, L137}

\bibitem[\protect\citeauthoryear{{Mu{\~n}oz}, {Carlin}, {Frinchaboy},
  {Nidever}, {Majewski}  \& {Patterson}}{{Mu{\~n}oz} et~al.}{2006}]{Munoz2006b}
{Mu{\~n}oz} R.~R.,  {Carlin} J.~L.,  {Frinchaboy} P.~M.,  {Nidever} D.~L.,
  {Majewski} S.~R.,   {Patterson} R.~J.,  2006, \mn@doi [\apjl]
  {10.1086/508685}, \href
  {https://ui.adsabs.harvard.edu/abs/2006ApJ...650L..51M} {650, L51}

\bibitem[\protect\citeauthoryear{{Mu{\~n}oz}, {Cote}, {Santana}, {Geha},
  {Simon}, {Oyarzun}, {Stetson}  \& {Djorgovski}}{{Mu{\~n}oz}
  et~al.}{2018}]{Munoz18}
{Mu{\~n}oz} R.~R.,  {Cote} P.,  {Santana} F.~A.,  {Geha} M.,  {Simon} J.~D.,
  {Oyarzun} G.~A.,  {Stetson} P.~B.,   {Djorgovski} S.~G.,  2018, arXiv
  e-prints, \href {https://ui.adsabs.harvard.edu/abs/2018arXiv180606891M} {p.
  arXiv:1806.06891}

\bibitem[\protect\citeauthoryear{{Muratov}, {Kere{\v{s}}},
  {Faucher-Gigu{\`e}re}, {Hopkins}, {Quataert}  \& {Murray}}{{Muratov}
  et~al.}{2015}]{Muratov2015}
{Muratov} A.~L.,  {Kere{\v{s}}} D.,  {Faucher-Gigu{\`e}re} C.-A.,  {Hopkins}
  P.~F.,  {Quataert} E.,   {Murray} N.,  2015, \mn@doi [\mnras]
  {10.1093/mnras/stv2126}, \href
  {https://ui.adsabs.harvard.edu/abs/2015MNRAS.454.2691M} {454, 2691}

\bibitem[\protect\citeauthoryear{{O{\~n}orbe}, {Boylan-Kolchin}, {Bullock},
  {Hopkins}, {Kere{\v{s}}}, {Faucher-Gigu{\`e}re}, {Quataert}  \&
  {Murray}}{{O{\~n}orbe} et~al.}{2015}]{Onorbe2015}
{O{\~n}orbe} J.,  {Boylan-Kolchin} M.,  {Bullock} J.~S.,  {Hopkins} P.~F.,
  {Kere{\v{s}}} D.,  {Faucher-Gigu{\`e}re} C.-A.,  {Quataert} E.,   {Murray}
  N.,  2015, \mn@doi [\mnras] {10.1093/mnras/stv2072}, \href
  {https://ui.adsabs.harvard.edu/abs/2015MNRAS.454.2092O} {454, 2092}

\bibitem[\protect\citeauthoryear{{Pontzen} \& {Governato}}{{Pontzen} \&
  {Governato}}{2012}]{Pontzen2012}
{Pontzen} A.,  {Governato} F.,  2012, \mn@doi [\mnras]
  {10.1111/j.1365-2966.2012.20571.x}, \href
  {https://ui.adsabs.harvard.edu/abs/2012MNRAS.421.3464P} {421, 3464}

\bibitem[\protect\citeauthoryear{{Read} \& {Gilmore}}{{Read} \&
  {Gilmore}}{2005}]{Read2005}
{Read} J.~I.,  {Gilmore} G.,  2005, \mn@doi [\mnras]
  {10.1111/j.1365-2966.2004.08424.x}, \href
  {https://ui.adsabs.harvard.edu/abs/2005MNRAS.356..107R} {356, 107}

\bibitem[\protect\citeauthoryear{{Revaz} \& {Jablonka}}{{Revaz} \&
  {Jablonka}}{2018}]{Revaz2018}
{Revaz} Y.,  {Jablonka} P.,  2018, \mn@doi [\aap]
  {10.1051/0004-6361/201832669}, \href
  {https://ui.adsabs.harvard.edu/abs/2018A&A...616A..96R} {616, A96}

\bibitem[\protect\citeauthoryear{{Saviane}, {Held}, {Momany}  \&
  {Rizzi}}{{Saviane} et~al.}{2001}]{Saviane01}
{Saviane} I.,  {Held} E.~V.,  {Momany} Y.,   {Rizzi} L.,  2001, \memsai, \href
  {https://ui.adsabs.harvard.edu/abs/2001MmSAI..72..773S} {72, 773}

\bibitem[\protect\citeauthoryear{{Schroyen}, {de Rijcke}, {Valcke},
  {Cloet-Osselaer}  \& {Dejonghe}}{{Schroyen} et~al.}{2011}]{Schroyen11}
{Schroyen} J.,  {de Rijcke} S.,  {Valcke} S.,  {Cloet-Osselaer} A.,
  {Dejonghe} H.,  2011, \mn@doi [\mnras] {10.1111/j.1365-2966.2011.19083.x},
  \href {https://ui.adsabs.harvard.edu/abs/2011MNRAS.416..601S} {416, 601}

\bibitem[\protect\citeauthoryear{{Simon}}{{Simon}}{2019}]{Simon19}
{Simon} J.~D.,  2019, \mn@doi [\araa] {10.1146/annurev-astro-091918-104453},
  \href {https://ui.adsabs.harvard.edu/abs/2019ARA&A..57..375S} {57, 375}

\bibitem[\protect\citeauthoryear{{Sparre}, {Hayward}, {Feldmann},
  {Faucher-Gigu{\`e}re}, {Muratov}, {Kere{\v{s}}}  \& {Hopkins}}{{Sparre}
  et~al.}{2017}]{Sparre2017}
{Sparre} M.,  {Hayward} C.~C.,  {Feldmann} R.,  {Faucher-Gigu{\`e}re} C.-A.,
  {Muratov} A.~L.,  {Kere{\v{s}}} D.,   {Hopkins} P.~F.,  2017, \mn@doi
  [\mnras] {10.1093/mnras/stw3011}, \href
  {https://ui.adsabs.harvard.edu/abs/2017MNRAS.466...88S} {466, 88}

\bibitem[\protect\citeauthoryear{{Stewart}, {Brooks}, {Bullock}, {Maller},
  {Diemand}, {Wadsley}  \& {Moustakas}}{{Stewart} et~al.}{2013}]{Stewart2013}
{Stewart} K.~R.,  {Brooks} A.~M.,  {Bullock} J.~S.,  {Maller} A.~H.,  {Diemand}
  J.,  {Wadsley} J.,   {Moustakas} L.~A.,  2013, \mn@doi [\apj]
  {10.1088/0004-637X/769/1/74}, \href
  {https://ui.adsabs.harvard.edu/abs/2013ApJ...769...74S} {769, 74}

\bibitem[\protect\citeauthoryear{{Tolstoy} et~al.,}{{Tolstoy}
  et~al.}{2004}]{Tolstoy04}
{Tolstoy} E.,  et~al., 2004, \mn@doi [\apjl] {10.1086/427388}, \href
  {https://ui.adsabs.harvard.edu/abs/2004ApJ...617L.119T} {617, L119}

\bibitem[\protect\citeauthoryear{{Tolstoy}, {Hill}  \& {Tosi}}{{Tolstoy}
  et~al.}{2009}]{Tolstoy09}
{Tolstoy} E.,  {Hill} V.,   {Tosi} M.,  2009, \mn@doi [\araa]
  {10.1146/annurev-astro-082708-101650}, \href
  {https://ui.adsabs.harvard.edu/abs/2009ARA&A..47..371T} {47, 371}

\bibitem[\protect\citeauthoryear{Van Der~Walt, Colbert  \& Varoquaux}{Van
  Der~Walt et~al.}{2011}]{vanderWalt2011}
Van Der~Walt S.,  Colbert S.~C.,   Varoquaux G.,  2011, Computing in Science \&
  Engineering, 13, 22

\bibitem[\protect\citeauthoryear{{Vargas}, {Geha}  \& {Tollerud}}{{Vargas}
  et~al.}{2014}]{Vargas14}
{Vargas} L.~C.,  {Geha} M.~C.,   {Tollerud} E.~J.,  2014, \mn@doi [\apj]
  {10.1088/0004-637X/790/1/73}, \href
  {https://ui.adsabs.harvard.edu/abs/2014ApJ...790...73V} {790, 73}

\bibitem[\protect\citeauthoryear{{Walker}, {Mateo}, {Olszewski}, {Bernstein},
  {Wang}  \& {Woodroofe}}{{Walker} et~al.}{2006}]{Walker2006}
{Walker} M.~G.,  {Mateo} M.,  {Olszewski} E.~W.,  {Bernstein} R.,  {Wang} X.,
  {Woodroofe} M.,  2006, \mn@doi [\aj] {10.1086/500193}, \href
  {https://ui.adsabs.harvard.edu/abs/2006AJ....131.2114W} {131, 2114}

\bibitem[\protect\citeauthoryear{{Walker}, {Mateo}, {Olszewski}, {Gnedin},
  {Wang}, {Sen}  \& {Woodroofe}}{{Walker} et~al.}{2007}]{Walker2007}
{Walker} M.~G.,  {Mateo} M.,  {Olszewski} E.~W.,  {Gnedin} O.~Y.,  {Wang} X.,
  {Sen} B.,   {Woodroofe} M.,  2007, \mn@doi [\apjl] {10.1086/521998}, \href
  {https://ui.adsabs.harvard.edu/abs/2007ApJ...667L..53W} {667, L53}

\bibitem[\protect\citeauthoryear{{Weisz}, {Dolphin}, {Skillman}, {Holtzman},
  {Gilbert}, {Dalcanton}  \& {Williams}}{{Weisz} et~al.}{2014a}]{Weisz2014a}
{Weisz} D.~R.,  {Dolphin} A.~E.,  {Skillman} E.~D.,  {Holtzman} J.,  {Gilbert}
  K.~M.,  {Dalcanton} J.~J.,   {Williams} B.~F.,  2014a, \mn@doi [\apj]
  {10.1088/0004-637X/789/2/147}, \href
  {https://ui.adsabs.harvard.edu/abs/2014ApJ...789..147W} {789, 147}

\bibitem[\protect\citeauthoryear{{Weisz}, {Dolphin}, {Skillman}, {Holtzman},
  {Gilbert}, {Dalcanton}  \& {Williams}}{{Weisz} et~al.}{2014b}]{Weisz2014b}
{Weisz} D.~R.,  {Dolphin} A.~E.,  {Skillman} E.~D.,  {Holtzman} J.,  {Gilbert}
  K.~M.,  {Dalcanton} J.~J.,   {Williams} B.~F.,  2014b, \mn@doi [\apj]
  {10.1088/0004-637X/789/2/148}, \href
  {https://ui.adsabs.harvard.edu/abs/2014ApJ...789..148W} {789, 148}

\bibitem[\protect\citeauthoryear{{Wheeler} et~al.,}{{Wheeler}
  et~al.}{2019}]{Wheeler2019}
{Wheeler} C.,  et~al., 2019, \mn@doi [\mnras] {10.1093/mnras/stz2887}, \href
  {https://ui.adsabs.harvard.edu/abs/2019MNRAS.490.4447W} {490, 4447}

\bibitem[\protect\citeauthoryear{{Wilkinson}, {Kleyna}, {Evans}, {Gilmore},
  {Irwin}  \& {Grebel}}{{Wilkinson} et~al.}{2004}]{Wilkinson2004}
{Wilkinson} M.~I.,  {Kleyna} J.~T.,  {Evans} N.~W.,  {Gilmore} G.~F.,  {Irwin}
  M.~J.,   {Grebel} E.~K.,  2004, \mn@doi [\apjl] {10.1086/423619}, \href
  {https://ui.adsabs.harvard.edu/abs/2004ApJ...611L..21W} {611, L21}

\makeatother
\end{thebibliography}

%\begin{thebibliography}{99}
%\end{thebibliography}

%%%%%%%%%%%%%%%%%%%%%%%%%%%%%%%%%%%%%%%%%%%%%%%%%%

%%%%%%%%%%%%%%%%% APPENDICES %%%%%%%%%%%%%%%%%%%%%

\appendix

\section{Simulated Galaxy Properties \& Ancillary Results}
We list the properties for all of the simulated galaxies in our sample in Table \ref{table:galaxy_stats}. These properties include the galaxies', masses, $\rm V_{max}$ values, mean projected 2D half mass radii over 100 different viewing angles, median stellar ages, and their projected stellar metallicity gradients.

\begin{figure}
	\includegraphics[width=\columnwidth, height=0.32
	\textheight,, trim = 0 0 0 0]{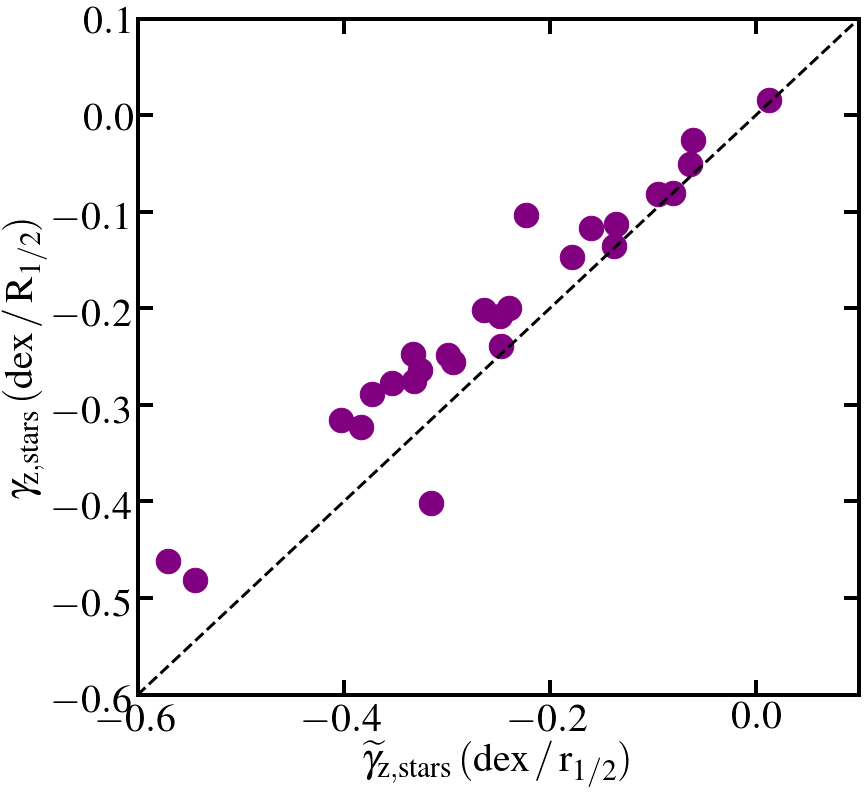}
	\centering
	\caption[migration all]{--- \textit{\textbf{2D versus 3D gradients}.} The relationship between the median projected 2D and 3D stellar metallicity gradients in our sample. A 1-to-1 line is depicted by the black, dashed line.}
	\label{fig:gradgrad}
\end{figure}

Figure \ref{fig:gradgrad} shows the relationship between the 2D projected stellar metallicity gradients ($\gamma_z$) and the 3D stellar metallicity gradients ($\widetilde{\gamma}_{z}$) for all of the galaxies in our sample. The black, dashed line is a 1-to-1 line. Whilst there is a strong relationship between $\gamma_z$ and $\widetilde{\gamma}_{z}$, they do not follow the 1-to-1 line because the values of $R_{1/2}$ (2D) and $r_{1/2}$ (3D) are slightly different.

\begin{figure*}
    \includegraphics[width=0.45\textwidth, trim = 0 0 0 0]{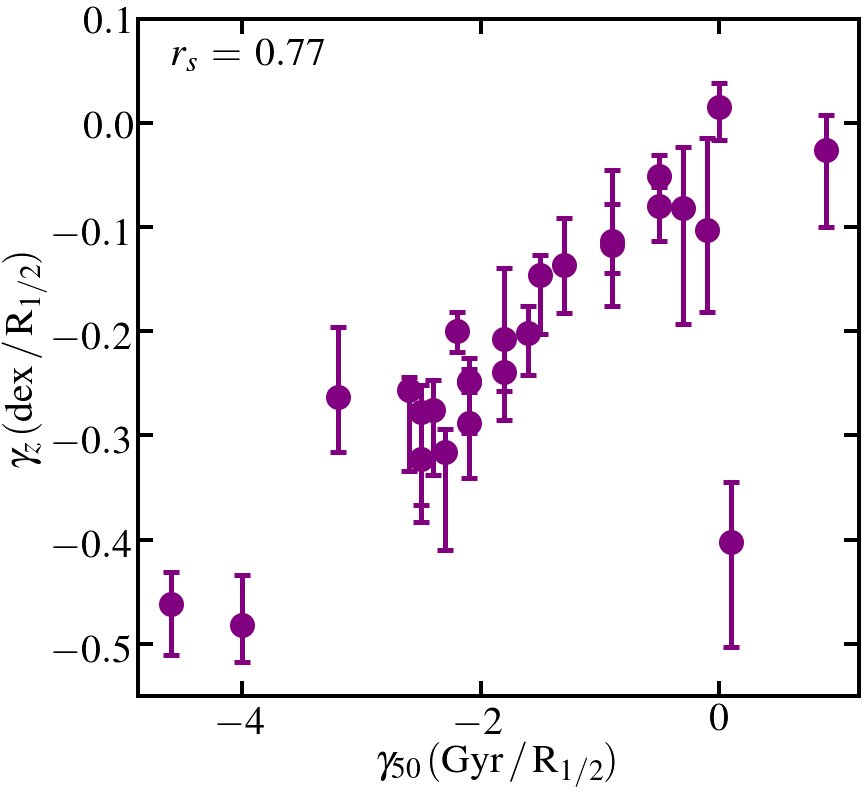}
    \includegraphics[width=0.45\textwidth, trim = 0 0 0 0]{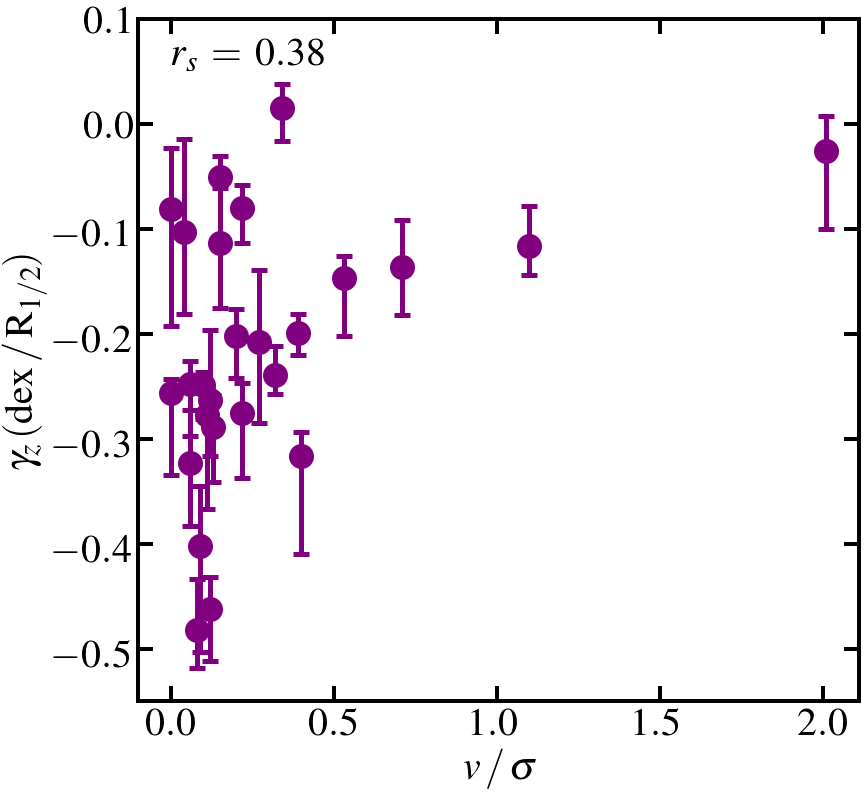}
	\includegraphics[width=0.45\textwidth, trim = 0 0 0 0]{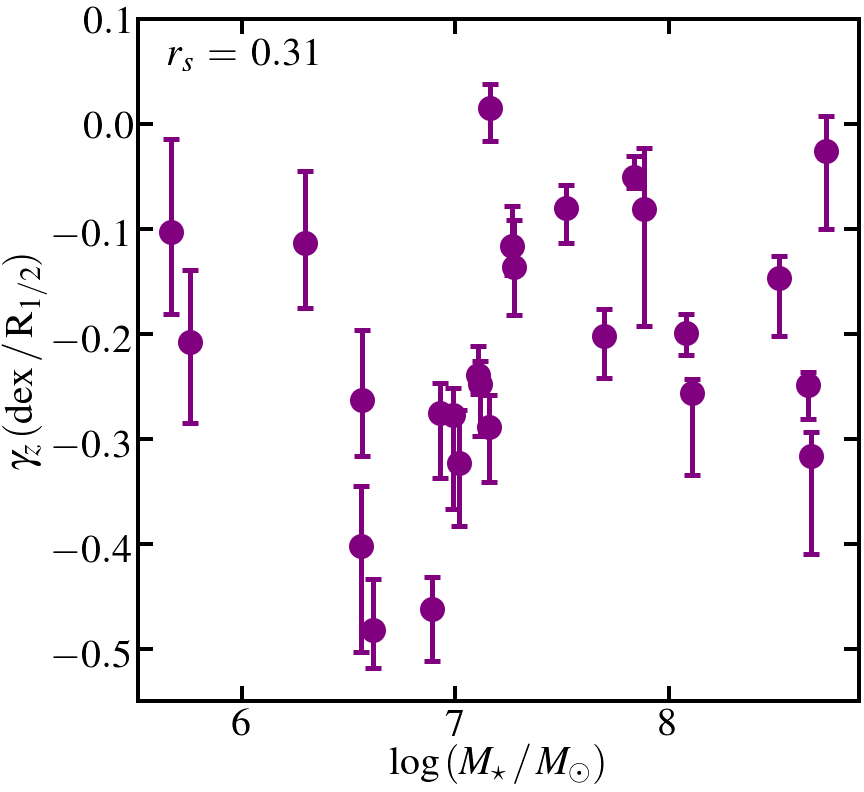}
	\includegraphics[width=0.45\textwidth, trim = 0 0 0 0]{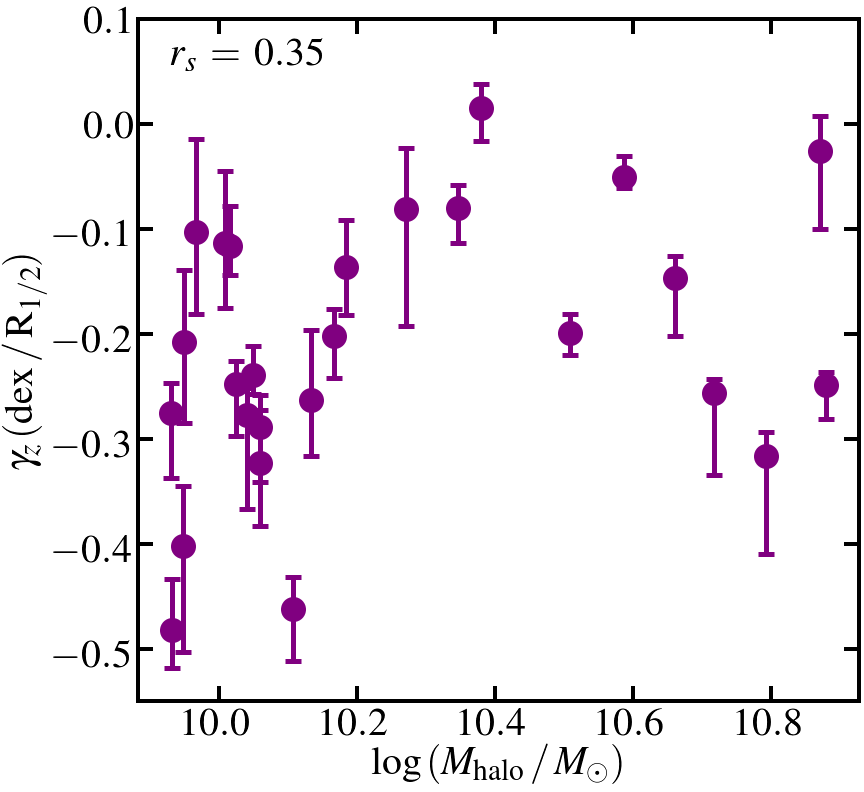}
    \includegraphics[width=0.45\textwidth, trim = 0 0 0 0]{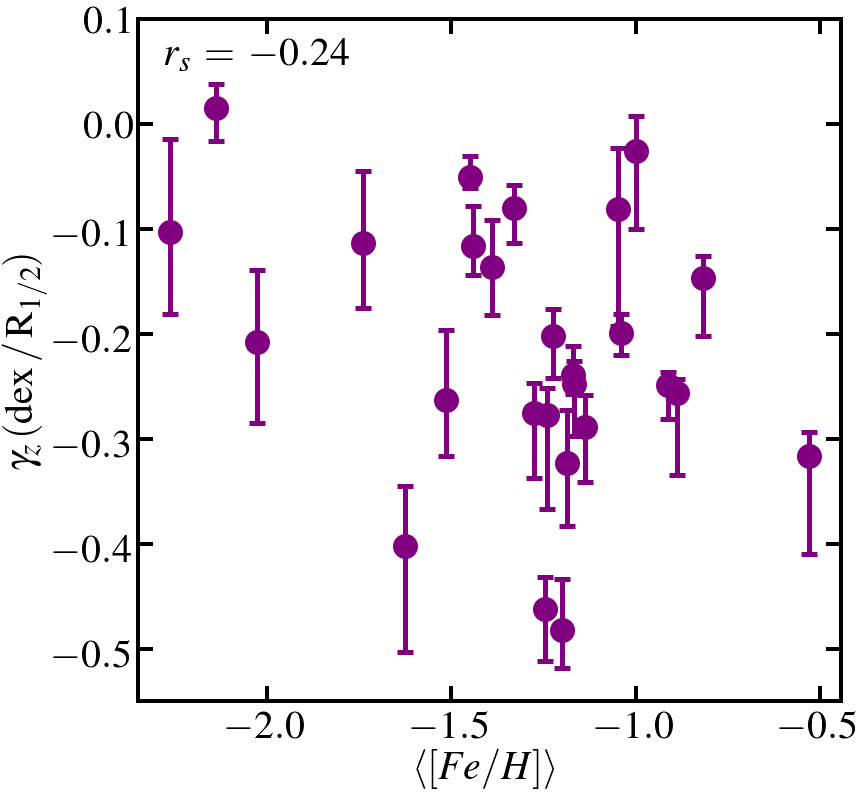}
	\includegraphics[width=0.45\textwidth, trim = 0 0 0 0]{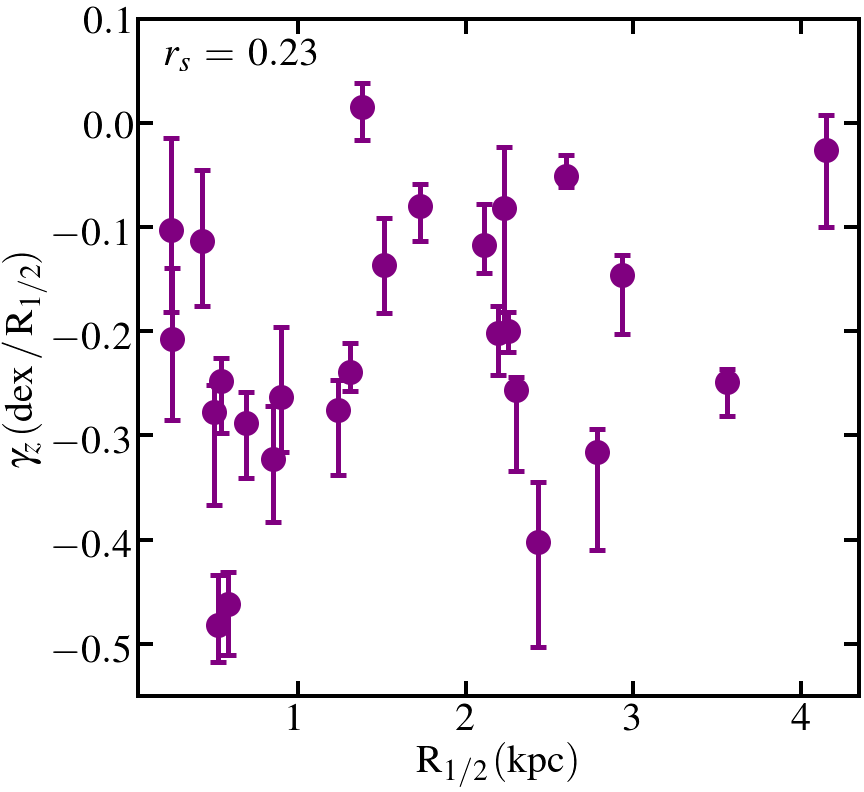}
	\centering
	\caption[appendix fig]{ --- \textbf{Stellar metallicity gradient strength versus several galaxy properties.} We test the relationships between the galaxies' stellar metallicity gradient strengths and their: age gradient strength \citep[$\gamma_{50}$, as presented in][]{Graus19}, $v/\sigma$, stellar and dark matter halo mass, median [Fe/H], and half mean, projected 2D half-mass radius over 100 random projections, $R_{1/2}$. We provide the spearman coefficients, $r_{s}$, at the top left of each panel as a measure of how well each relationship is correlated.}
	\label{fig:grad_other}
\end{figure*}

Figure \ref{fig:grad_other} depicts the relationship (or lack thereof) between the stellar metallicity gradients of the galaxies in our sample and other galaxy properties. The top left panel shows the relationship between the metallicity gradients calculated in this work and the age gradients calculated in \citet{Graus19}. As expected, due to the stellar age-metallicity relation, there is a positive correlation between metallicity gradients and age gradients such that galaxies with strong/weak metallicity gradients also tend to have strong/weak age gradients. 

Recent studies have indicated that, within the Local Group, rotation-supported dIrrs exhibit weaker gradients whilst dispersion-supported dSphs exhibit stronger gradients. This suggests that a galaxy's gradient strength is somehow tied to it's $v/\sigma$ value \citep{Schroyen11, Leaman13, Kacharov17}. The top right panel shows that within our simulated sample there is no correlation between the two. However, it is important to note that most observed dIrrs are more rotation-supported than the simulated galaxies in our sample. The next four panels illustrate that the metallicity gradients in our sample are independent of stellar mass, halo mass, $V_{\rm max}$, and half-mass radius.

%%%%%%%%%%%%%%%%%%%% Table
%\begin{comment}
\begin{table*}
  \caption{Properties of the simulated galaxies at $\it{z}$ = 0 used in this work: (1) stellar mass within 10\% of the virial radius of the halo,  (2) dark matter halo mass, (3) maximum circular velocity, (4) mean 2D half-mass radius over all viewing angles; (5) lookback time to the formation of 50\% of stars within 10\% of the virial radius of the halo; (6) median stellar metallicity gradients over all projections including the maximum and minimum gradients as error bars. Note that we present the same values for (2), (3), and (4) as \citet{Graus19}. We present m10xg and m10xh in bold as they are an example of systems with comparable stellar masses but different stellar metallicity gradients.}

\centering % used for centering table
%\begin{tabularx}{\textwidth}{lccccc}
\begin{tabularx}{\textwidth}{lXXXXXX}
\hline
\hline  %inserts double horizontal lines
  $\rm Halo$ &       $M_{\rm star}$ &        $M_{\rm halo}$ &      $\rm V_{max}$ &    $\rm R_{1/2}$ & \tfifty & $\gamma_{z}$ \\
  $\rm Name$ &  $\rm [\msun]$ & $\rm [\msun]$ & [$\rm km\, s^{-1}$] & [kpc] & [Gyr] & [dex/\rm $R_{1/2}$]\\
  
\hline 
\citet{Graus19} & (1) & (2) & (3) & (4) & (5) & (6) \\

\hline 
\vspace{0.01cm}\\
m10xa     &   $7.64 \times 10^{7}$   &   $1.87 \times 10^{10}$  &  45.26  &  2.23  &  6.08  &  $-0.08^{+0.07}_{-0.15}$ \vspace{0.2cm}\\
m10xb     &  $3.29 \times 10^{7}$   &   $2.22 \times 10^{10}$  &  42.78  &  1.73  &  4.23  &  $-0.09^{+0.04}_{-0.05}$ \vspace{0.2cm}\\
m10xc     &   $1.19 \times 10^{8}$   &   $3.22 \times 10^{10}$  &  48.31  &  2.25  &  6.55  &  $-0.20^{+0.03}_{-0.06}$ \vspace{0.2cm}\\
$\rm m10xc\_A$  &   $8.46 \times 10^{6}$   &   $8.52 \times 10^{9}$   &  35.03  &  1.24  &  10.89 &  $-0.34^{+0.05}_{-0.06}$ \vspace{0.2cm}\\
m10xd     &   $6.81 \times 10^{7}$   &   $3.86 \times 10^{10}$  &  53.51  &  2.60  &  4.04  &  $-0.07^{+0.02}_{-0.01}$ \vspace{0.2cm}\\
$\rm m10xd\_A$  &   $1.44 \times 10^{7}$   &   $2.40 \times 10^{10}$  &  38.52  &  1.38  &  1.63  &   $0.01^{+0.04}_{-0.03}$ \vspace{0.2cm}\\
m10xe     &   $3.26 \times 10^{8}$   &   $4.57 \times 10^{10}$  &  56.17  &  2.93  &  6.13  &  $-0.17^{+0.02}_{-0.08}$ \vspace{0.2cm}\\
$\rm m10xe\_A$  &   $3.64 \times 10^{6}$   &   $1.36 \times 10^{10}$  &  35.74  &  0.90  &  8.50  &  $-0.32^{+0.09}_{-0.10}$ \vspace{0.2cm}\\
$\rm m10xe\_B$  &   $1.28 \times 10^{7}$   &   $1.12 \times 10^{10}$  &  38.15  &  1.31  &  8.75  &  $-0.28^{+0.03}_{-0.04}$ \vspace{0.2cm}\\
$\rm m10xe\_C$  &   $1.84 \times 10^{7}$   &   $1.04 \times 10^{10}$  &  34.43  &  2.11  &  7.08  &  $-0.13^{+0.05}_{-0.05}$ \vspace{0.2cm}\\
$\rm m10xe\_D$  &   $3.61 \times 10^{6}$   &   $8.88 \times 10^{9}$   &  34.13  &  2.43  &  9.62  &  $-0.50^{+0.07}_{-0.13}$ \vspace{0.2cm}\\
m10xf     &   $1.28 \times 10^{8}$   &   $5.21 \times 10^{10}$  &  58.47  &  2.30  &  7.38  &  $-0.28^{+0.01}_{-0.08}$ \vspace{0.2cm}\\
\textbf{m10xg }    &   $4.61 \times 10^{8}$   &   $6.20 \times 10^{10}$  &  65.75  &  2.78  &  7.59  &  $-0.33^{+0.04}_{-0.13}$ \vspace{0.2cm}\\
$\rm m10xg\_A$  &   $1.88 \times 10^{7}$   &   $1.53 \times 10^{10}$  &  40.31  &  1.51  &  5.11  &  $-0.13^{+0.05}_{-0.10}$ \vspace{0.2cm}\\
\textbf{m10xh }    &   $5.40 \times 10^{8}$   &   $7.44 \times 10^{10}$  &  68.10  &  4.15  &  3.65  &  $-0.03^{+0.04}_{-0.06}$ \vspace{0.2cm}\\
$\rm m10xh\_A$  &   $4.97 \times 10^{7}$   &   $1.47 \times 10^{10}$  &  38.80  &  2.19  &  5.68  &  $-0.17^{+0.04}_{-0.05}$ \vspace{0.2cm}\\
m10xi     &   $4.48 \times 10^{8}$   &   $7.58 \times 10^{10}$  &  64.35  &  3.56  &  6.03  &  $-0.31^{+0.02}_{-0.04}$ \vspace{0.2cm}\\

\hline

\citet{Fitts17} &  &  &  & \\

\hline
\vspace{0.01cm}\\
m10b  &   $4.65 \times 10^{5}$  &  $9.29 \times 10^{9}$   &  31.51  &  0.24  &   2.54  & $-0.13^{+0.10}_{-0.12}$ \vspace{0.2cm}\\
m10c  &	  $5.75 \times 10^{5}$  &  $8.92 \times 10^{9}$   &  31.40  &  0.25  &   4.07  &  $-0.23^{+0.10}_{-0.12}$ \vspace{0.2cm}\\
m10e  &	  $1.98 \times 10^{6}$  &  $1.02 \times 10^{10}$  &  31.44  &  0.43  &   5.63  &  $-0.14^{+0.05}_{-0.07}$ \vspace{0.2cm}\\
m10f  &   $4.11 \times 10^{6}$  &  $8.56 \times 10^{9}$   &  35.66  &  0.52  &  11.96  &  $-0.59^{+0.05}_{-0.06}$ \vspace{0.2cm}\\
m10h  &   $7.80 \times 10^{6}$  &  $1.28 \times 10^{10}$  &  37.98  &  0.58  &  11.64  &  $-0.54^{+0.04}_{-0.07}$ \vspace{0.2cm}\\
m10j  &   $9.74 \times 10^{6}$  &  $1.10 \times 10^{10}$  &  37.98  &  0.50  &  11.51  &  $-0.30^{+0.04}_{-0.11}$ \vspace{0.2cm}\\
m10k  &   $1.04 \times 10^{7}$  &  $1.15 \times 10^{10}$  &  38.22  &  0.85  &  10.74  &  $-0.33^{+0.06}_{-0.10}$ \vspace{0.2cm}\\
m10l  &   $1.30 \times 10^{7}$  &  $1.06 \times 10^{10}$  &  37.62  &  0.54  &  10.76  &  $-0.27^{+0.04}_{-0.07}$ \vspace{0.2cm}\\
m10m  &   $1.44 \times 10^{7}$  &  $1.15 \times 10^{10}$  &  38.51  &  0.69  &   9.86  &  $-0.29^{+0.04}_{-0.08}$ \vspace{0.2cm}\\
\hline
\hline
\label{table:galaxy_stats}
\end{tabularx}
\end{table*}

%\end{comment}

%%%%%%%%%%%%%%%%%%%%%%%%%%%%%%%%%%%%%%%%%%%%%%%%%%

% Don't change these lines
\bsp	% typesetting comment
\label{lastpage}
\end{document}